\documentclass[fleqn,usenatbib]{mnras}

\usepackage[T1]{fontenc}

\DeclareRobustCommand{\VAN}[3]{#2}
\let\VANthebibliography\thebibliography
\def\thebibliography{\DeclareRobustCommand{\VAN}[3]{##3}\VANthebibliography}

\usepackage{graphicx}	
\usepackage{amsmath}	
\usepackage{amssymb}	
\usepackage{float}
\usepackage{setspace} 
\usepackage{threeparttable}
\usepackage{booktabs}
\usepackage{newtxtext,newtxmath}

\title[Investigating an iron spread in NGC 362]{Investigating a Predicted Metallicity [Fe/H] Variation in the Type II Globular Cluster NGC 362}

\author[C. Vargas et al.]{
C. Vargas$^{1}$,
S. Villanova$^{1}$,
D. Geisler$^{1,2,3}$,
C. Mu\~noz$^{1,2}$,
L. Monaco$^{4}$,
J. O'Connell$^1$.
Ata Sarajedini$^5$
\\
\\
$^{1}$Departamento de Astronom\'ia, Casilla 160-C, Universidad de Concepci\'on, Concepci\'on, Chile.\\
$^{2}$Instituto de Investigaci\'on Multidisciplinario en Ciencia y Tecnolog\'ia, 
Universidad de La Serena. Avenida Ra\'ul Bitr\'an S/N, La Serena, Chile.\\
$^{3}$Departamento de  Astronom\'ia, Facultad de Ciencias, Universidad de La Serena. Av. Juan Cisternas 1200, La Serena, Chile.\\
$^{4}$Departamento de Ciencias Físicas, Universidad Andrés Bello, Fernández Concha 700, Las Condes Santiago, Chile.\\
$^5$Department of Physics, Florida Atlantic University, 777 Glades Rd., Boca Raton, FL 33431, USA
}

\date{Accepted XXX. Received YYY; in original form ZZZ}

\pubyear{2015}

\begin{document}
\label{firstpage}
\pagerange{\pageref{firstpage}--\pageref{lastpage}}
\maketitle

\begin{abstract}
NGC 362 is a non-common Type II Galactic globular cluster, showing a complex pseudo two-color diagram or 'chromosome map'. The clear separation of its stellar populations in the color-magnitude diagram and the distribution of the giant stars in the chromosome map strongly suggests that NGC 362 could host stars with both cluster-nominal as well as enhanced heavy-element abundances, and one of them could be iron. However, despite previous spectroscopic observations of NGC 362, no such iron variation has been detected.
Our main goal is to confirm or disprove this result by searching for any internal variation of [Fe/H] which would give us insight into the formation and evolution of this interesting globular cluster. In this paper, we present the abundance analysis for a sample of 11 red giant branch members based on high-resolution and high S/N spectra obtained with the MIKE echelle spectrograph mounted at the Magellan-Clay telescope. HST and GAIA photometry and astrometry has been used to determine atmospheric parameters and membership. We obtained T$_{\text{eff}}$, log(g) and v$_{\text{t}}$ for our target stars and measured the mean iron content of the sample and its dispersion with three different methods, which lead to [Fe/H]$_1$=-1.10$\pm0.02$, [Fe/H]$_2$=-1.09$\pm0.01$ and [Fe/H]$_3$=-1.10$\pm0.01$, while the internal dispersion turned out to be $\sigma_{[\text{Fe/H}]_1}$=0.06$\pm0.01$, $\sigma_{[\text{Fe/H}]_2}$=0.03$\pm0.01$ and $\sigma_{[\text{Fe/H}]_3}$=0.05$\pm0.01$ respectively. The error analysis gives an internal dispersion due to observational error of 0.05 dex. Comparing the observed dispersion with the internal errors, we conclude that NGC 362 does not show any trace of an internal iron spread.
\end{abstract}




\section{Introduction}

Globular Clusters (GCs) are perfect laboratories for studying a wide variety of fundamental astrophysical problems since they are among the oldest known objects in the universe, they are bright and present in large numbers in the Galaxy. GCs were traditionally considered Simple Stellar Populations (SSPs), with all of the stars having the same age, distance and initial chemical composition. However it is now well known that almost all GCs host Multiple Stellar Populations (MSPs), indicating an intrinsic range in present composition, which is evidenced by both spectroscopic and photometric methods. A comprehensive review of GCs abundance anomalies has been extensively investigated by \cite{Gratton2012,Charbonnel2016,Bastian2018} and \citet{Gratton2019} and we refer the interested readers to these works for further details about the MSPs phenomenon.

Through high-resolution spectra in particular we can detect intracluster chemical variations in light elements, most notably C, N, O, Na, Mg and Al \citep{Carretta}. These star-to-star variations display (anti)correlations, best evidenced by a Na-O anticorrelation, and point to the likely presence of different generations of stars, formed in more than a single burst. The first generation (1G) of stars have primordial chemical composition, where stars are O rich and Na poor, the same as typical field stars of the same metallicity, while the next generation(s) (NG) of stars are born in a medium chemically enriched by the earlier generation(s), resulting in enhanced He, N, Al and Na abundances and depleted C, O and Mg. This chemical signature suggests that multiple generations of stars are formed from gas polluted by the products of hot proton-capture processes which take place in the interiors of the 1G stars, which eventually reach the surface and then escape via stellar winds. The GC potential well is able to retain this gas and eventually leads to the subsequent formation of chemically enriched stars \citep{DErcole}. Several mechanisms and polluter candidates have been proposed as responsible for the observed nucleosynthesis in GCs, such as intermediate-mass ($\sim$4-8$\text{M}_{\odot}$) asymptotic giant branch (AGB) stars \citep{Ventura2001,D'Antona2016}, fast rotating massive main sequence (MS) stars \citep{Meynet2006,Decressin}, interacting massive binaries \citep{deMink:2009yg}, supermassive (M$>$10$^4\text{M}_{\odot}$) stars \citep{Denissenkov2014}, and the accretion model \citep{Batian2013}. However, the possible polluters cannot be fully established yet and no current scenario is able to completely reproduce the complex behavior of MSPs \citep{Renzini}.

The iron abundance homogeneity found in most GCs studied to date tells us that supernovae events cannot be responsible for the chemically enriched medium from which the NG formed. Although GCs are generally able to retain the gas from stellar winds, leading to MSPs, they are not massive enough to retain the Fe-peak elements produced by supernovae ejecta \citep{Willman}. There are some exceptions, such as NGC 5139 ($\omega$ Centauri) and Terzan 5, that display an intrinsic dispersion in Fe of $\sim$1 dex. Nonetheless, these objects are now classified as the remnant core of a tidally disrupted dwarf galaxy accreted by the Milky Way \citep{Majewski} and as a Galactic bulge primordial building block \citep{Ferraro2009}, respectively. We can also find a few traditional GCs with highly debated iron spreads in the literature, such as NGC 6656 (M22) \citep{Marino,Mucciarelli2015}. However the origin of these unusual clusters is even less clear than that of the more common ones that only show light element, and no heavy element, variations.

\begin{figure}
\centering
\includegraphics[width=3.6in,height=3.0in]{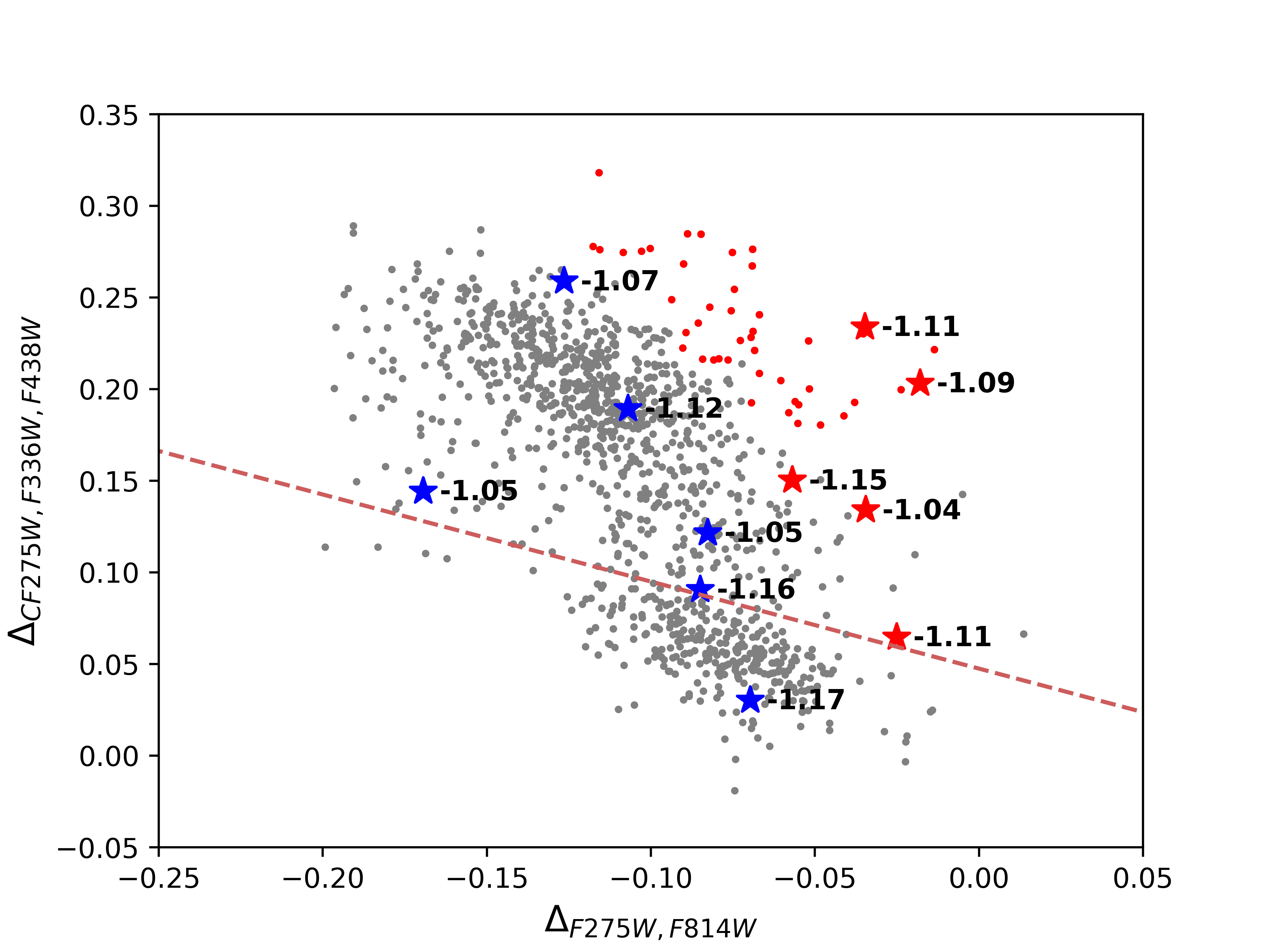}
\caption{Chromosome map of NGC 362. Our targets are represented as red (anomalous) and blue (normal) filled stars, as plotted in Figure \ref{CMD}. The magenta dashed line is used to separate 1G from 2G stars and the red points indicate the red-RGB stars according to the definition of \citet{Milone17}. Each target's spectroscopically-derived metallicity is noted.}
\label{ChromosomeMap1.png}
\end{figure}

The complexity of MSPs can also be very well characterized photometrically using an appropriate combination of ultraviolet (UV) and optical Hubble Space Telescope (HST) photometry \citep{Milone2012,Milone2013,Piotto2013}. The F275W, F336W and F438W filters, also known as the ’Magic Trio’, of the HST UV and Visual Channel of the Wide Field Camera 3 (WFC3/UVIS) are very sensitive to OH, NH, CN and CH molecular bands and hence do an excellent job of separating MSPs in appropriate color-color and color-magnitude diagrams (CMDs) \citep{2015AJ....149...91P}. \citet{Milone17} found a method to maximize this separation through the construction of the ’Chromosome Map’; this pseudo two-color diagram makes use of the magic trio and of the F814W filter to produce a significant split of different stellar populations \citep{Milone2015a, Milone2015b}. By using high-precision photometry of red giant branch (RGB) stars in 57 GCs from The HST UV Legacy Survey of Galactic GCs \citep{2015AJ....149...91P, Milone17}, Milone et al. found a surprising variety of MSP behavior, with each of them displaying distinct distributions. Based on their chromosome maps, the authors were able to identify certain generic patterns and define two classes of GCs: Type I clusters show the typical bimodality where the 1G and NG stars are almost always clearly separated and no other strong pattern emerges, whereas Type II GCs display more complex chromosome maps, with 1G and/or NG sequences that appear to be split, with an additional distribution of very red stars. A detailed analysis of the HST photometry of Type II clusters exposed that sub-giant branches (SGBs) are also split in UV-optical CMDs, with a (less populated) fainter SGB joining onto a redder RGB, while the SGBs of Type I clusters split only in the chromosome map. Further spectroscopic analysis revealed that stars in the faint SGB and red RGB belonging to well-studied Type II clusters are enhanced in iron, global C+N+O content, and generally, in s-process elements compared to stars on the normal, bluer RGB, whereas Type I clusters host populations that show only light element variations \citep{Milone17}. Hence, these Type II GCs display intrinsic variations in heavy elements, in particular Fe, meaning they possess a metallicity spread. 

\begin{figure}
\centering
\includegraphics[width=3.6in,height=3.0in]{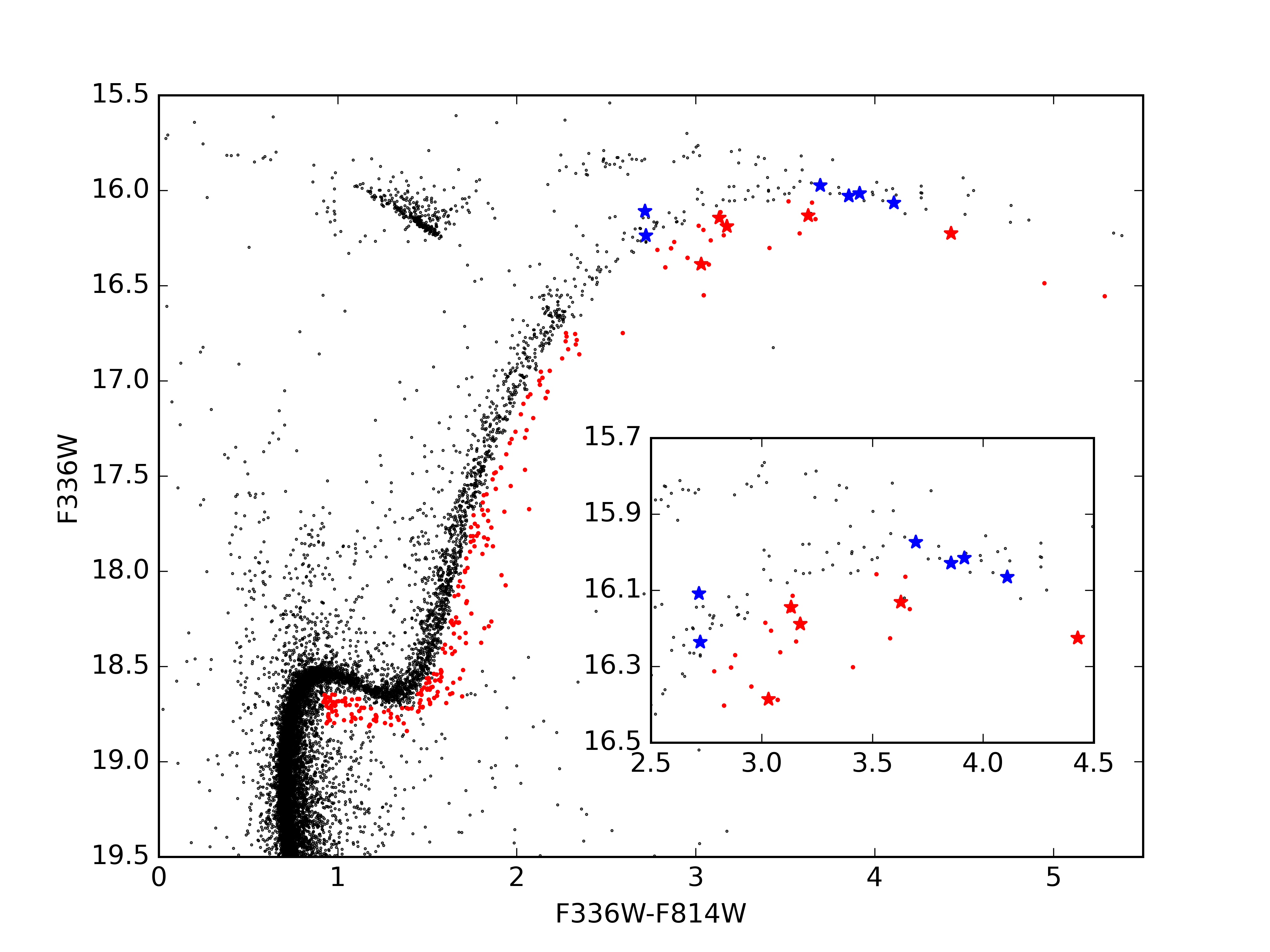}
  \caption{The CMD of NGC 362 from the HUGS catalogue \citep{2018MNRAS.481.3382N,2015AJ....149...91P} with the eleven observed RGB members indicated as blue and red filled stars. Blue targets are presumed to have normal metallicities, while red targets should have higher metallicities according to the \citet{Milone17} scenario. Red points represent the redder RGB and fainter SGB.}
 \label{CMD}
 \end{figure}

\begin{table*}
\centering
\scriptsize
\caption{Coordinates, proper motions, V, GAIA and HST photometry, and radial velocities for the target stars. Star $\#$12 has been added for completeness, but has been disregarded in the following analysis. The V band was calculated using F438W, F606W and F814W filters.}
\begin{tabular}{@{}ccccccccccccccc@{}}
\toprule
ID & RA                & DEC                & pmRA  & pmDEC  & V      & G      & B$_p$  & R$_p$  & F275W  & F336W  & F438W  & F606W  & F814W  & RV$_{helio}$ \\
 &{\scriptsize{}(°)} & {\scriptsize{}(°)}   & {\scriptsize{}(mas/yr)} &{\scriptsize{}(mas/yr)} & {\scriptsize{}(mag)} &{\scriptsize{}(mag)} &{\scriptsize{}(mag)} &{\scriptsize{}(mag)} &{\scriptsize{}(mag)} &{\scriptsize{}(mag)} &{\scriptsize{}(mag)} &{\scriptsize{}(mag)} &{\scriptsize{}(mag)} &{\scriptsize{}(km/s)}\\
 \midrule
11 & 15.77331529496288 & -70.84201303750254 & 6.505 & -2.207 & 13.62  & 13.375 & 13.872 & 12.497 & 18.765 & 16.131 & 15.020 & 13.415 & 12.503 & 225.91     \\
12 & 15.80827162077069 & -70.84453738823032 & 6.573 & -1.783 & 13.64  & 13.422 & 13.491 & 12.332 & 18.718 & 16.140 & 15.064 & 13.490 & 12.600 & 215.34     \\
13 & 15.83355100868370 & -70.83207022949118 & 6.687 & -2.463 & 14.07  & 13.841 & 14.385 & 13.094 & 18.591 & 16.188 & 15.319 & 13.883 & 13.014 & 230.30     \\
14 & 15.84985498855671 & -70.84020798422353 & 6.606 & -2.279 & 14.09  & 13.856 & 14.363 & 13.077 & 18.621 & 16.144 & 15.313 & 13.893 & 13.012 & 222.99     \\
15 & 15.80960339082632 & -70.86619177758531 & 6.685 & -2.597 & 14.35  & 14.147 & 14.616 & 13.398 & 18.527 & 16.386 & 15.610 & 14.172 & 13.356 & 230.55     \\
16 & 15.75481176194973 & -70.83878187983893 & 6.966 & -2.696 & 14.50  & 14.300 & 14.766 & 13.570 & 18.389 & 16.236 & 15.621 & 14.324 & 13.514 & 226.95     \\
17 & 15.79282209680344 & -70.83773058066414 & 6.606 & -2.335 & 13.05  & 12.736 & 13.402 & 11.896 & 19.236 & 16.225 & 14.667 & 12.815 & 11.798 & 225.40     \\
18 & 15.84874591604511 & -70.86713946413691 & 6.478 & -2.139 & 13.19  & 12.897 & 13.584 & 12.096 & 18.787 & 16.065 & 14.697 & 12.961 & 11.957 & 218.33     \\
19 & 15.86879292727561 & -70.85824653897895 & 6.665 & -2.629 & 13.31  & 13.018 & 13.672 & 12.227 & 18.862 & 16.015 & 14.804 & 13.085 & 12.100 & 215.49     \\
20 & 15.78729221784991 & -70.86387947858030 & 6.775 & -2.665 & 13.35  & 13.071 & 13.686 & 12.258 & 18.754 & 16.028 & 14.769 & 13.134 & 12.173 & 211.90     \\
21 & 15.79577687002238 & -70.86514268003519 & 6.886 & -2.667 & 13.44  & 13.167 & 13.794 & 12.378 & 18.830 & 15.973 & 14.847 & 13.226 & 12.277 & 215.72     \\
22 & 15.82674179749346 & -70.86406858433082 & 6.728 & -2.645 & 14.40  & 14.205 & 14.696 & 13.480 & 18.169 & 16.108 & 15.553 & 14.223 & 13.391 & 217.04     \\ \bottomrule
\label{Tabla1}
\end{tabular}
\end{table*}

\begin{figure}
\centering
\includegraphics[width=3.6in,height=3.0in]{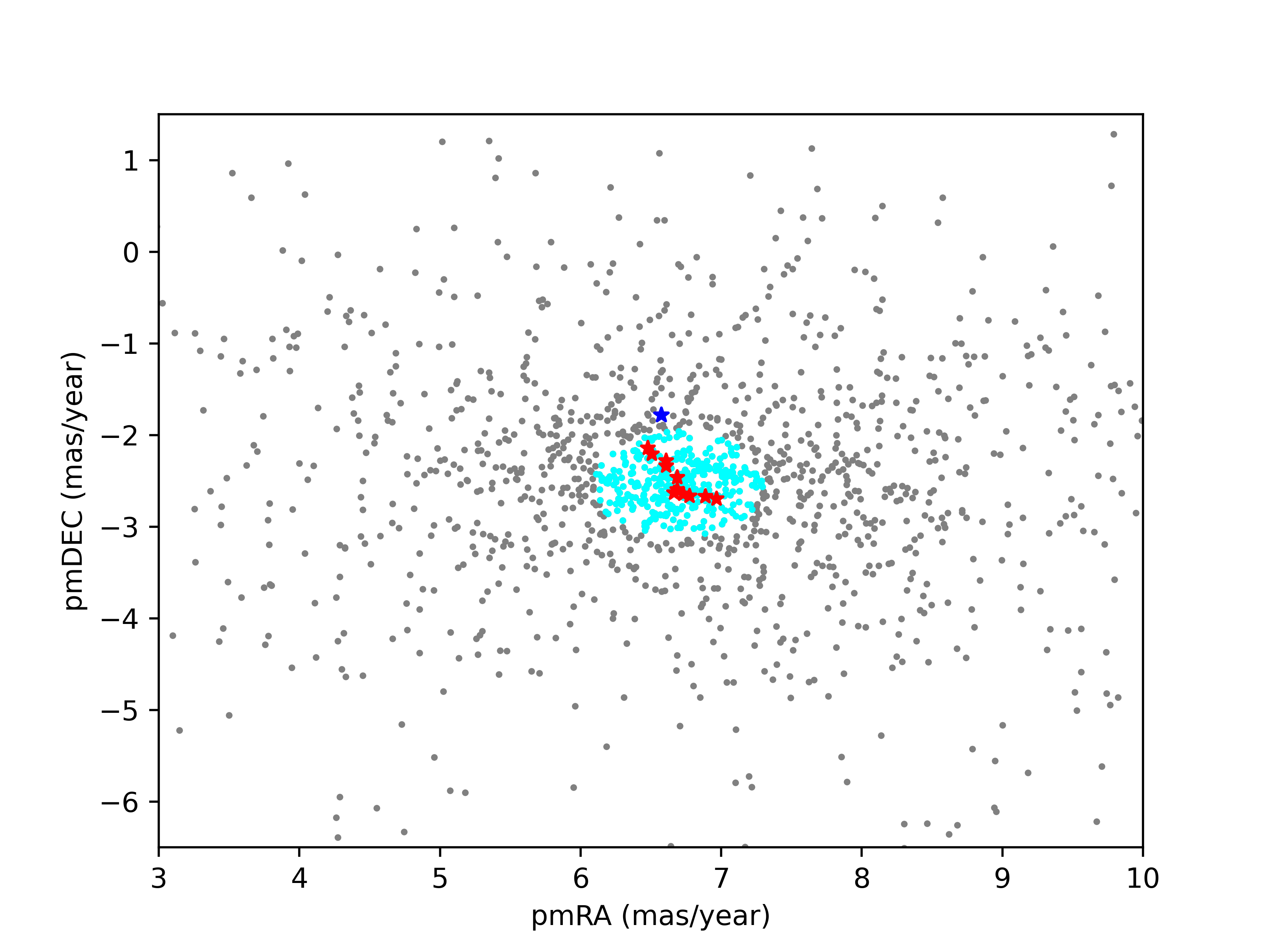}
\caption{Proper motions in the field of NGC 362. Presumed possible cluster members within a 0.6 mas/year circle are indicated with cyan dots, while our targets are represented as red filled stars. The blue filled star represents the rejected star number 12 (See Section \ref{Observations and data reduction}). Data was taken from GAIA eDR3 survey \citep{GAIA2020}.}
\label{Proper_Motions.png}
\end{figure}

\begin{figure*}
\centering
\includegraphics[width=6.2in,height=5.0in]{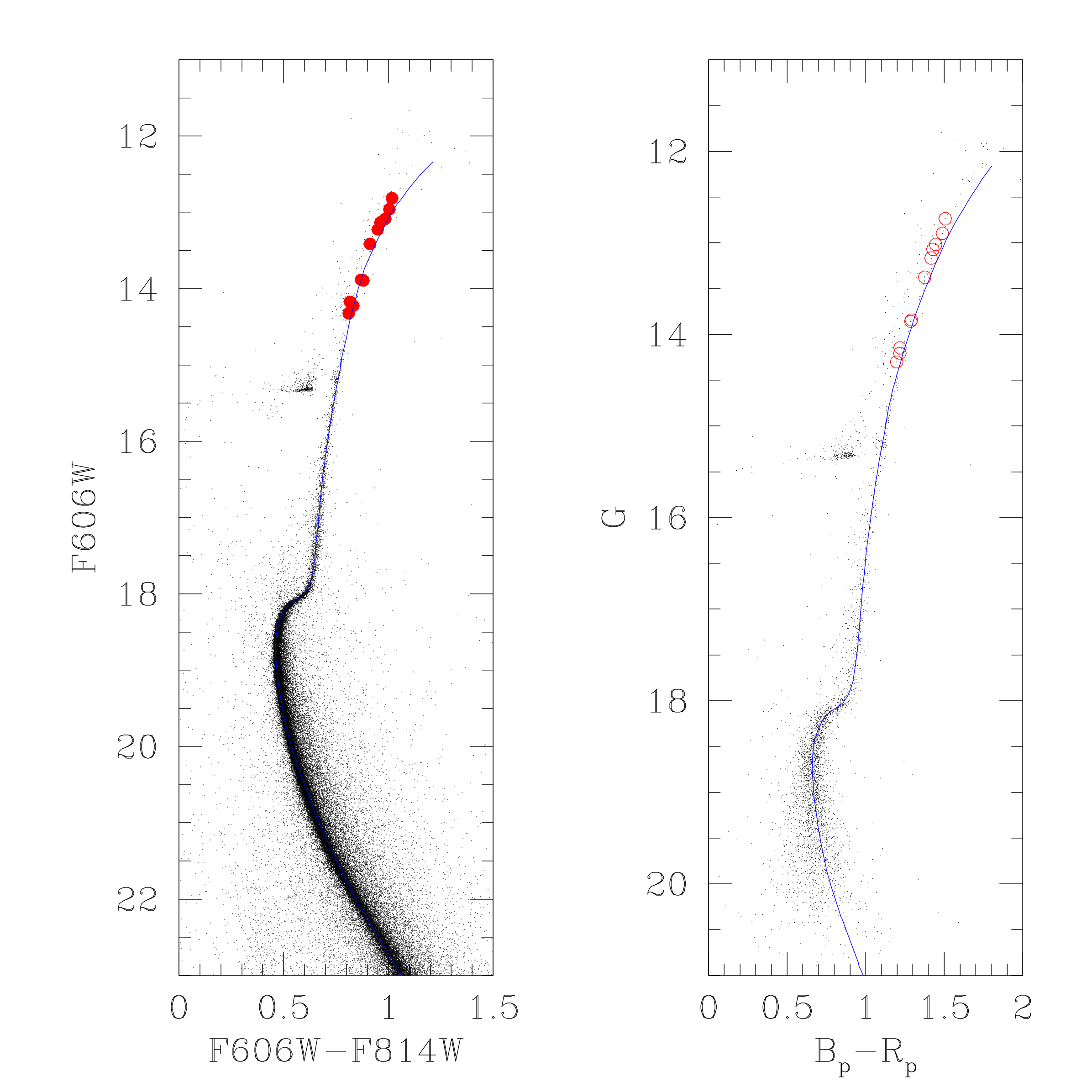}
\caption{CMDs of NGC 362 from HST and GAIA photometry with isochrones obtained from PARSEC database \citep{2012MNRAS.427..127B}. Red symbols represent the observed RGB stars.}
\label{Teff_logg_GAIA_HST.png}
\end{figure*} 

\begin{figure}
\centering
\includegraphics[width=3.5in,height=3.0in]{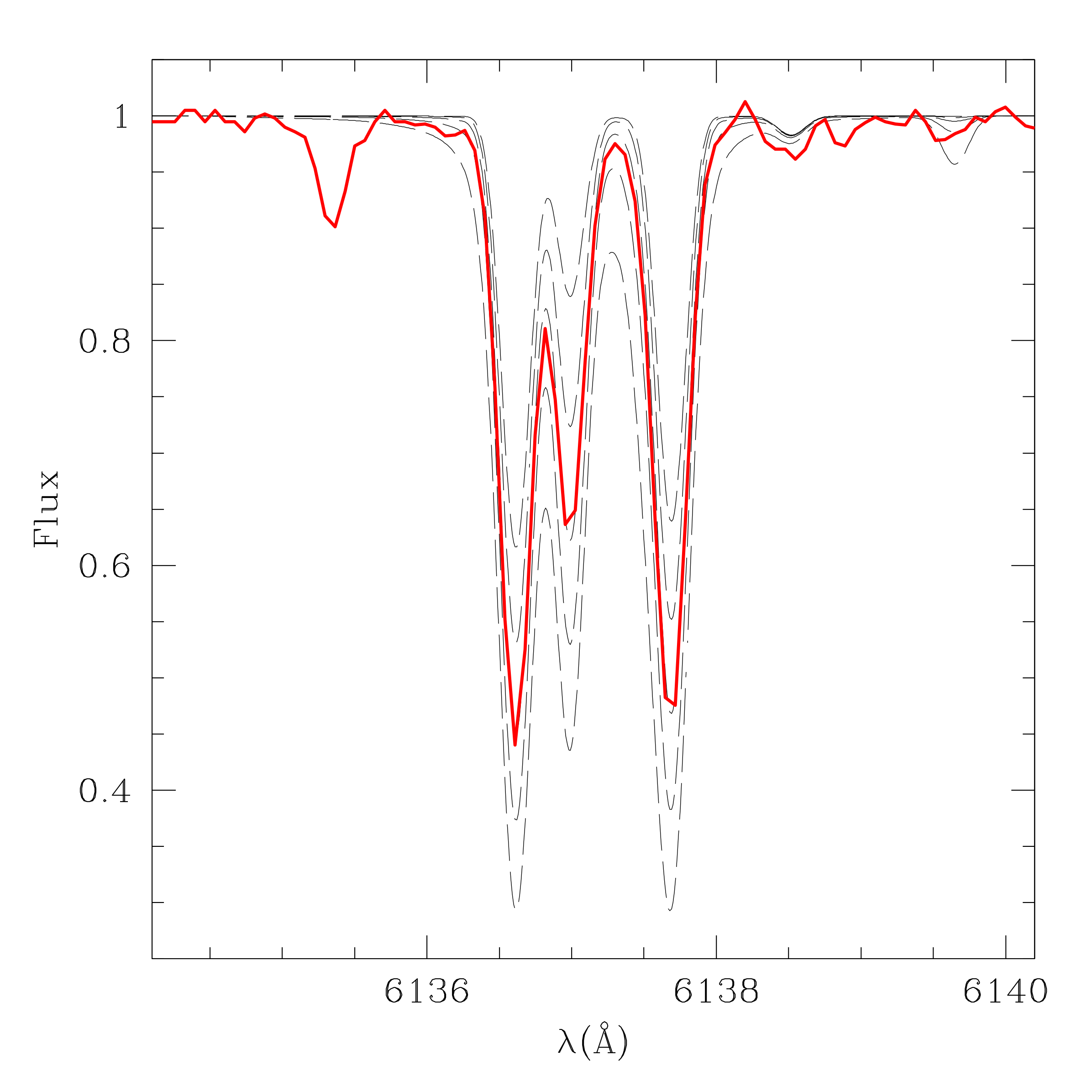}
\caption{Example of the strong Fe band at 6137 \AA\ used as Fe indicator. The star used for this plot was target $\#15$, whose derived [Fe/H] ratio is $-1.05$ dex. The observed spectrum is represented by the red line and overplotted are five synthetic spectra that we used to determine Fe abundances. Synthetic spectra were calculated with an abundance of -0.50, -0.25, 0.00, +0.25 and +0.50 dex with respect to the [Fe/H] content we estimated for the star.}
\label{Fe_6127.png}
\end{figure}

Chromosome maps have proven to be an efficient tool to identify candidate GCs with internal variations of heavy elements, which are until now very rare and their study essential for understanding GC formation and evolution. All the known clusters with a significant iron spread like $\omega$ Centauri belong to Type II, however there are several clusters in this category that are currently not known to have Fe abundance spreads. In a recent study, \citet{Meszaros20} analyzed abundances of RGB stars in 31 GCs from high-resolution spectra, including a sample of 6 Type II GCs, i.e., NGC 5139, NGC 7089 (M2), NGC 6388, NGC 1851, M22 and NGC 362. However, despite their Type II classification, a significant iron spread could only by detected in the very complex cluster $\omega$ Centauri. On the other hand, \citet{Carretta2021} derived a clear-cut answer on the intrinsic metallicity distribution in NGC 6388 by using high-resolution spectra of giant stars in order to obtain a chemical characterisation of MSPs hosted in this massive Type II GC. The results show that no indication of an iron spread is present within the cluster. Indeed, they concluded that either an intrinsic metallicity dispersion is not a mandatory requisite to explain the Type II classification or NGC 6388 does not belong to this group. \citet{Carretta10,Carretta2011} discussed the case of the Type II GC NGC 1851, where a small but detectable metallicity spread was found by using spectroscopic data; nonetheless, these results could not be confirmed by either \citet{Marino2019} nor \citet{Meszaros20}, whereas \citet{Tautvaisien2022} reported a difference of 0.07 dex in the mean metallicity values between the metal-rich and metal-poor populations. The existence of intrinsic [Fe/H] variations in NGC 6656 (M22) and M2 seems to be more ambiguous. \citet{Yong2014} found an extremely wide metallicity distribution in M2, which was later confirmed by \citet{Marino2019}, yet a previous study using EQW measurements by \citet{Lardo16} concluded that this cluster is composed by a dominant population ($\sim$99\%) homogeneous in Fe and a minority component ($\sim$1\%) enriched in iron. Similarly, \citet{Marino2009,Marino} suggested that there may indeed be star-to-star iron variations in M22, however neither \citet{Mucciarelli2015} nor \citet{Meszaros20} could corroborate those results later. In addition to the GCs listed as a Type II in \citet{Milone17}, we also added the metal-poor GC NGC 7078 (M15) to the discussion, which was considered by \citet{Nardiello2018} as a Type II from the analysis of its chromosome map, and that according to \citet{Carretta2009} does not show evidence of an internal metallicity dispersion (see also \citet{Carretta,Carretta2009a}). Therefore, there is a fraction of Type II GCs where no [Fe/H] variations were detected. To date, metallicity spreads have been confirmed to be present in five Type II GCs, including as previously mentioned, $\omega$ Centauri \citep{Johnson2010,Gratton2011,Marino2019,Meszaros20}, NGC 6934 \citep{Marino18,Marino2021}, NGC 5286 \citep{Marino2015,Marino2019}, NGC 1261 \citep{Marino2021,Munoz21} and NGC 6715 (M54) \citep{Carretta2010b,Marino2019}. So, in general, the evidence suggests that the group of the so called Type II GCs seems to be heterogeneous. In this paper we present a detailed chemical study of the Type II GC NGC 362 in order to search for any intrinsic Fe variation, which will allow us to address the problem of discrepant fractions of Type II GCs where no metallicity dispersions were found. The existence of a possible metallicity spread in NGC 362 has been previously studied from both photometric and spectroscopic data (e.g. \citet{Shetrone2000, Worley10,Carretta13,Bailin19,Marino2019,Meszaros20}). However, our study is the first one that explicitly select purported normal and anomalous stars (see Section \ref{Target selection}) a priori to compare metallicities between them. The evident split of 1G and 2G sequences with a significant distribution of red stars in the Chromosome Map (see Figure \ref{ChromosomeMap1.png}), leading to its Type II classification, strongly argues that NGC 362 should have an intrinsic metallicity spread, including iron. \citet{Milone17} estimate that this cluster could host up to $\sim$7.5\% of stars having an enhanced metal content higher than the main population and, if so, the Fe spread could be of the order of 0.1-0.2 dex. Usually metallicity variations are associated with massive GCs; however, NGC 362 has an intermediate mass of $3.5\text{x}10^{5} \text{M}_{\odot}$ \citep{2018MNRAS.478.1520B}. Therefore, if an Fe spread is confirmed, it will help us to constrain formation scenarios, as well as the lower limit to the mass of a GC required to retain supernovae ejecta. For this purpose, we collected high-resolution spectra for 11 RGB member stars taken with the MIKE spectrograph, selected to cover a wide range in the chromosome map and CMDs to include stars predicted to have a metallicity spread. This data has been complemented with HST and GAIA photometry to fully characterize the atmospheric parameters and confirm or reject the hypothesis of the presence of an iron spread, as the chromosome map suggests.

The paper is organized as follows: the target selection is presented in Section \ref{Target selection}; observations and data reduction are in Section \ref{Observations and data reduction}; the derivation of the atmospheric parameters and the comparison of the different techniques used are in Section \ref{Atmospheric parameters determination}. In Section \ref{Results} we report our iron content results with their respective dispersions; and finally the summary and conclusion are in Section \ref{Summary and conclusions}.

\section{Target Selection} \label{Target selection}

We first reproduced the chromosome map of NGC 362\footnote{Coordinates and $\Delta_{\text{F275W,F814W}}$, $\Delta_{\text{CF275W,F336W,F438W}}$ values of all the stars in the Chromosome Map will be available for online download.}, nevertheless we did not apply corrections for differential reddening, as this is beyond our main goal and not crucial to investigate the existence of an iron spread in the cluster. To do this, we used the F275W, F336W, F438W and F814W magnitudes from The HST UV Legacy Survey \citep{2018MNRAS.481.3382N} and formed the two pseudo-colors $\Delta$F275W,F814W and $\Delta$CF275W,F336W,F438W. The chromosome map is plotted in Figure \ref{ChromosomeMap1.png} and clearly shows the dominant presence of typical 1G and second generation (2G) stars (black points). We also plotted the F336W vs. (F336-F814W) CMD (see Figure \ref{CMD}). This plot shows a main as well as a loosely populated redder RGB that connects to a fainter SGB. Unfortunately, neither \citet{2015AJ....149...91P} nor \citet{Milone17} published the coordinates of the different stellar populations and red stars for NGC 362, so we selected as red stars all those that are located to the right side of the main RGB (and are therefore redder) as well as those that are below the SGB (which are therefore fainter) and marked these same stars as red in the chromosome map (Figure \ref{ChromosomeMap1.png}). These red stars fall to redder $\Delta$F275W,F814W colors of the main locus of 2G stars. These outliers (hereby anomalous stars) were identified by \citet{Milone17} and led to their classification of NGC 362 as a Type II GC. These anomalous stars presumably have a higher metallicity than the bulk of the stars that fall along the nominal 1G and 2G sequences (hereby normal stars), according to their scenario. Thus, we need to observe as many of these anomalous stars as possible, along with a similar number of normal (metallicity) 1G/2G stars, in order to optimize the possibility of detecting any putative metallicity spread between them \citep{Marino18}. Our final targets included the brightest  anomalous stars from the chromosome map along with bright normal stars chosen from the CMD which lay along the bluer/brighter RGB, in order to maximize the possibility to detect any variation in iron spread. 

\begin{table*}
\centering
\caption{Columns 2-9: T$_{\text{eff}}$ and log(g) obtained through isochrone fitting using GAIA and HST data. Column 10: log(g) obtained from the canonical equation. Columns 11-12: T$_{\text{eff}}$ and log(g) used to calculate v$_\text{t}$ and [Fe/H]. Columns 13-14: Microturbulence and metallicities obtained assuming T$_{\text{eff}}$ and log(g) from columns 11 and 12, and using FeI lines from the spectra. Columns 14-15: Derived [FeII/H] abundances and classification of the stars, 'n' for normal and 'a' for anomalous.}
\begin{tabular}{@{}cccccccccccccccc@{}}
\toprule
ID &
  \tiny{T$_{\text{GAIA$_{\text{m}11}$}}$} &
  \tiny{g$_{\text{GAIA$_{\text{m}11}$}}$} &
  \tiny{T$_{\text{GAIA$_{\text{m}09}$}}$} &
  \tiny{g$_{\text{GAIA$_{\text{m}09}$}}$} &
  \tiny{T$_{\text{HST$_{\text{m}11}$}}$} &
  \tiny{g$_{\text{HST$_{\text{m}11}$}}$} &
  \tiny{T$_{\text{HST$_{\text{m}09}$}}$} &
  \tiny{g$_{\text{HST$_{\text{m}09}$}}$} &
  log(g)$_{\text{phot}}$ &
  T$_{\text{eff}}$ &
  log(g) &
  v$_\text{t}$ &
 [Fe/H] &
 [FeII/H] &
 Class \\ \midrule
11 & 4405 & 1.28 & 4306 & 1.27 & 4415 & 1.30 & 4303 & 1.26 & 1.25 & 4303 & 1.26 & 1.61 & -1.11 & -1.14 & a \\
13 & 4558 & 1.56 & 4465 & 1.55 & 4573 & 1.58 & 4470 & 1.56 & 1.55 & 4470 & 1.56 & 1.40 & -1.02 & -1.03 & a \\
14 & 4563 & 1.56 & 4470 & 1.56 & 4576 & 1.59 & 4473 & 1.56 & 1.56 & 4473 & 1.56 & 1.48 & -1.11 & -1.20 & a \\
15 & 4652 & 1.73 & 4563 & 1.72 & 4662 & 1.74 & 4564 & 1.73 & 1.70 & 4564 & 1.73 & 1.44 & -1.05 & -1.02 & a \\
16 & 4696 & 1.81 & 4609 & 1.81 & 4706 & 1.83 & 4610 & 1.81 & 1.82 & 4610 & 1.81 & 1.32 & -1.07 & -0.99 & n \\
17 & 4176 & 0.87 & 4072 & 0.85 & 4187 & 0.89 & 4064 & 0.83 & 0.88 & 4064 & 0.83 & 1.69 & -1.05 & -1.04 & a \\
18 & 4236 & 0.98 & 4132 & 0.96 & 4246 & 1.00 & 4124 & 0.94 & 1.02 & 4124 & 0.94 & 1.51 & -1.07 & -1.15 & n \\
19 & 4281 & 1.06 & 4177 & 1.04 & 4294 & 1.08 & 4175 & 1.03 & 1.10 & 4175 & 1.03 & 1.59 & -1.18 & -1.25 & n \\
20 & 4299 & 1.09 & 4198 & 1.07 & 4312 & 1.12 & 4196 & 1.07 & 1.14 & 4196 & 1.07 & 1.57 & -1.17 & -1.10 & n \\
21 & 4333 & 1.15 & 4232 & 1.13 & 4347 & 1.18 & 4231 & 1.13 & 1.18 & 4231 & 1.13 & 1.54 & -1.20 & -1.14 & n \\
22 & 4669 & 1.76 & 4581 & 1.76 & 4677 & 1.77 & 4580 & 1.76 & 1.77 & 4580 & 1.76 & 1.39 & -1.07 & -1.06 & n \\ \bottomrule
\end{tabular}
\label{Tabla2}
\end{table*}

\begin{table}
\centering
\caption{Columns 2-3: T$_{\text{eff}}$ and log(g) from the previous table. Columns 4-5: Microturbulence obtained from the \citet{Mott2020} relation, and metallicity obtained using spectral synthesis of FeI lines and considering vt$_{\text{Mott}}$. Column 6: Microturbulence obtained from the \citet{2008A&A...490..625M} relation. Column 7: Classification of the stars.}
\begin{tabular}{@{}ccccccc@{}}
\toprule
ID & T$_{\text{eff}}$    & log(g)    & vt$_{\text{Mott}}$ & [Fe/H]    & vt$_{\text{Marino}}$ & Class \\ \midrule                        
11 & 4303 & 1.26 & 1.61       & -1.09 & 1.61    & a   \\                                 
13 & 4470 & 1.56 & 1.55       & -1.05 & 1.53    & a   \\                                 
14 & 4473 & 1.56 & 1.56       & -1.09 & 1.53    & a   \\                                 
15 & 4564 & 1.73 & 1.52       & -1.04 & 1.49    & a   \\                                 
16 & 4610 & 1.81 & 1.51       & -1.06 & 1.47    & n   \\                                 
17 & 4064 & 0.83 & 1.70       & -1.09 & 1.72    & a   \\                                 
18 & 4124 & 0.94 & 1.66       & -1.10 & 1.69    & n   \\                                 
19 & 4175 & 1.03 & 1.66       & -1.13 & 1.67    & n   \\                                 
20 & 4196 & 1.07 & 1.65       & -1.10 & 1.66    & n   \\                                 
21 & 4231 & 1.13 & 1.64       & -1.13 & 1.64    & n   \\                                 
22 & 4580 & 1.76 & 1.51       & -1.10 & 1.48    & n   \\ \bottomrule
\end{tabular}
\label{Tabla3}
\end{table}

\section{OBSERVATIONS AND DATA REDUCTION} \label{Observations and data reduction}

Twelve candidate NGC 362 members, six normal and six anomalous stars,  were observed during the nights of November 8th and 9th 2018 with the MIKE high-resolution spectrograph \citep{Bernstein} mounted at the Magellan-II/Clay telescope of the Las Campanas Observatory (Chile). The spectra cover the range 3350-5000 \AA\ and 4850-9400 \AA\ in the blue and the red arms, respectively. A 0.7$\arcsec$ slit was used for the observations, corresponding to a spectral resolution of R=53000 and R=42000 in the blue and the red arms, respectively. Stars were observed using either 2x900s (stars \#11, 12, 13, 17, 18, 19, 20) or 3x1200s (stars \#14, 15, 16, 21, 22) exposures. The signal-to-noise (S/N) is between 57 and 105 around 6000 \AA. Data were reduced using the CarPy pipeline\footnote{\url{https://code.obs.carnegiescience.edu/mike}} \citep{Kelson_2000,Kelson_2003} and included overscan subtraction, flat fielding, order extraction, wavelength calibration and sky subtraction. Milky flat fields were taken with the aid of a diffuser to correct for the pixel to pixel variations. Spectral rectification was performed using \textit{continuum} and \textit{sarith} tasks in IRAF \footnote{IRAF is distributed by the National Optical Astronomy Observatory, which is operated by the Association of Universities for Research in Astronomy, Inc., under cooperative agreement with the National Science Foundation.}. We also used the  \textit{scombine} package to combine all the spectra per star and the  \textit{fxcor} package to measure radial velocities using a synthetic spectrum as a template. Observed radial velocities were converted to the heliocentric system leading to a mean value of $RV_{helio} = 221.87 \pm 1.95 \text{km s}^{-1}$, which is in good agreement with \citet{1996AJ....112.1487H} and \citet{Baumgardt19} who found values of $RV_{H}=223.50\pm0.5\text{km s}^{-1}$ and $RV_{B}=223.26\pm 0.28\text{km s}^{-1}$ respectively. Thus, this is a first strong indication, apart from their location in  the various CMDs, that our targets are all indeed members of NGC 362.

In order to further confirm the membership of our targets, we plot the proper motions of our stars around the cluster, making use of GAIA eDR3 survey \citep{GAIA2020}. Figure \ref{Proper_Motions.png} shows in cyan the most likely cluster members while our targets are indicated with red and blue filled stars. We considered as likely members those stars within a radius of 0.6 mas/year around the center of the cluster. It is clear from this figure that one of our targets (star number 12, an assumed anomalous star, marked in blue) lies significantly further away from the cluster mean than our other targets. We rejected this star from further analysis as its proper motion differs by 3.3$\sigma$ from the mean of the other stars. Furthermore, this star is located in the central crowded region of the cluster and therefore has unreliable photometric and spectroscopic data. We could see this when we tried to plot its position on the CMD. According to the HST photometry star \#12 lies on the RGB, while according to the GAIA photometry it better matches the AGB. For all these reasons we decided not to consider star \#12 in our analysis so we ended up with a sample of 11 RGB stars (5 anomalous and 6 normal stars). The mean proper motion of these stars is:
$$\text{pm}_{\text{RA}}=6.69\pm0.14\text{mas/year}$$ $$\text{pm}_{\text{DEC}}=-2.48\pm0.20\text{mas/year}$$
This is in very good agreement with \citet{Vasiliev2021}, who obtained a mean value for the cluster of $\text{pm}_{\text{RA}}=6.69\pm 0.024\text{mas/year}$ and  $\text{pm}_{\text{DEC}}=-2.53\pm0.024\text{mas/year}$. Considering both radial velocity and  proper motion analysis, as well as position in the cluster and along the main cluster loci in the CMD, we conclude that our final sample of 11 stars are all bona fide cluster members.

Table \ref{Tabla1} lists the basic parameters of the eleven final members as well as the disregarded star $\#12$: ID, coordinates, proper motions, G, $\text{B}_{\text{p}}$, $\text{R}_{\text{p}}$ magnitudes from GAIA DR2, F275W, F336W, F438W, F606W, F814W magnitudes from HST photometry, and heliocentric radial velocities. The V band was calculated using the F438W, F606W and F814W filters \citep{Harris2018TransformationOH}.

\section{Atmospheric parameters determination} \label{Atmospheric parameters determination}

In order to obtain the atmospheric parameters T$_{\text{eff}}$, log(g), v$_{\text{t}}$ and [Fe/H], we used three different methods. In the first, we derived T$_{\text{eff}}$ and log(g) by comparing the F606W vs. (F606W-F814W) CMD from HST photometry and the G vs. (B$_{P}$-R$_{P}$) CMD from GAIA photometry with isochrones obtained from the PARSEC database \citep{2012MNRAS.427..127B}. The fit is shown in Figure \ref{Teff_logg_GAIA_HST.png}. We obtained the best fit for an age of 10.5 Gyrs in both cases. This value is in very good agreement with \citet{Mar09} which gives an age of 11.0-10.5 Gyrs. The reddenings and distance moduli we estimated are E(F606W-F814W)=0.02 and (m-M)$_{F606W}$=14.85 for the HST CMD, and E(B$_{P}$-R$_{P}$)=0.05 and (m-M)$_{G}$=14.85 for the GAIA CMD, which is in good agreement with \citet{harris2010new} who found values of E(B-V)=0.05 and (m-M)$_{V}$=14.83. The other key parameter to select an isochrone is the metallicity, but PARSEC isochrones use solar-scaled abundances while GCs are $\alpha$-enhanced. A way to overcome this discrepancy is to use a global metallicity [M/H] that is calculated taking both [Fe/H] and  $\alpha$-enhancement into account and using the equation from \cite{1993ApJ...414..580S}. However [Fe/H] and [$\alpha$/Fe] are obtained only when atmospheric parameters are known. For this reason, we had to make an initial guess for [Fe/H] and [$\alpha$/Fe] based on the literature and then make an initial estimation for T$_{\text{eff}}$ and log(g). Using these parameters we then obtained revised estimates for [Fe/H] and [$\alpha$/Fe] and iterated the process until convergence. Our final value for the mean cluster metallicity is [Fe/H]=-1.1 (see Section \ref{Results}) and using [$\alpha$/Fe]=+0.3 we get [M/H]=-0.9. Another issue we had to solve is the discrepancy between the color of the isochrones and the color of the stars in the upper RGB since Figure \ref{Teff_logg_GAIA_HST.png} shows that isochrones tend to be redder than real stars, especially when we consider the G vs. (B$_{P}$-R$_{P}$) CMD. This is not surprising since isochrones are transformed from the theoretical temperature-luminosity plane to the observational plane by synthetic temperature-color relations. These relations are calculated using synthetic spectra that poorly reproduce the spectra of the very cool upper RGB stars, especially in the blue-UV part. We found that the best way to overcome this problem is projecting horizontally the position of the star in the CMD until it intersects the isochrone and assuming its T$_{\text{eff}}$ and log(g) to be the temperature and gravity of the point of the isochrones that have the same F606W or G magnitude. We checked also the impact of the age on the T$_{\text{eff}}$ determination. We found that an age variation of $\pm$1 Gyr implies a T$_{\text{eff}}$ variation of few degrees. We conclude that age does not affect our results in a significant way. The results of this first method are reported in Table \ref{Tabla2}.

\begin{table}
\centering
\caption{Parameters and metallicities obtained spectroscopically using the FeI and FeII lines.}
\begin{tabular}{@{}cccccc@{}}
\toprule                                                                             
ID & T$_{\text{eff}}$ & log(g) & v$_\text{t}$ & [Fe/H]  & Class   \\ \midrule                                 
11 & 4297      & 1.00    & 1.60  & -1.15 & a \\                                          
13 & 4475      & 1.58    & 1.49  & -1.04 & a \\                                          
14 & 4479      & 1.48    & 1.49  & -1.11 & a \\                                          
15 & 4486      & 1.54    & 1.38  & -1.11 & a \\                                          
16 & 4637      & 1.69    & 1.35  & -1.05 & n \\                                          
17 & 4111      & 0.62    & 1.70  & -1.09 & a \\                                          
18 & 4209      & 0.79    & 1.54  & -1.07 & n \\                                          
19 & 4250      & 0.90    & 1.61  & -1.16 & n \\                                          
20 & 4292      & 1.04    & 1.58  & -1.12 & n \\                                  
21 & 4300      & 1.08    & 1.56  & -1.17 & n \\
22 & 4614      & 1.74    & 1.42  & -1.05 & n \\ \bottomrule
\end{tabular}
\label{Tabla4}
\end{table}

We check the gravities by the canonical equation:
$$ \log\left(\frac{g}{g_{\odot}}\right) =
         \log\left(\frac{M}{M_{\odot}}\right)
         + 4 \log\left(\frac{T_{\rm{eff}}}{T_{\odot}}\right)
         - \log\left(\frac{L}{L_{\odot}}\right) $$

where the mass M was assumed to be 0.9 M$_{\odot}$ from the isochrone fitting, and the luminosity L/L$_{\odot}$ was obtained from the absolute magnitude M$_{\rm V}$ assuming an apparent distance modulus of (m-M)$_{\rm V}$=14.83 from \citet{1996AJ....112.1487H}. The bolometric correction (BC) was derived by adopting the relation BC-T$_{\rm eff}$ from \citet{1999A&AS..140..261A}. Both distance modulus and BC are valid for the Johnson V band, while we have HST magnitudes. For this reason, we transformed F438W, F606W and F814W magnitudes into Johnson B,V and I magnitudes using the transformations from \citet{Harris2018TransformationOH}. The V magnitudes and gravities we obtained are reported in Table \ref{Tabla1} and Table \ref{Tabla2} respectively. Table \ref{Tabla2} shows that gravities from isochrones agree very well with those obtained from the canonical equation and also that T$_{\text{eff}}$ and log(g) obtained independently from HST and GAIA data match each other within the errors. We point out that some of our stars could be AGB since at the magnitude of our targets it can be difficult to distinguish between giants and asymptotic giants. If this was the case, we should use a mass 0.2 M$_\odot$ lower in the canonical equation. This would lower the gravity of 0.1 dex as given by the canonical equation, which has negligible impact on our final FeI abundances.

For the iron content determination we used the parameters we obtained from the HST data. We also have to determine microturbulence v$_{\text{t}}$ together with [Fe/H]. This determination relies on the abundances obtained from equivalent width (EQW) of FeI lines as measured from the spectra. EQWs were measured fitting a gaussian to unblended and well defined iron lines. The line-list we used is the same used in previous papers (e.g. \citet{Villanova2013}), so we refer to those articles for a detailed discussion about this point. We underline here that particular care was applied to the continuum determination. First of all, atmospheric models were calculated using the ATLAS9 code \citep{1970SAOSR.309.....K}, assuming our photometric estimations of T$_{\rm eff}$ and log(g) as an initial guess. We assume also v$_{\rm t}$=1.5 km/s and [Fe/H]=-1.1 as initial estimations. Then v$_{\rm t}$ was re-adjusted and new atmospheric models calculated in an interactive way in order to remove trends in reduced equivalent width versus abundance plots. About 100 FeI lines (depending on the S/N of the spectrum) were used for this purpose. The [Fe/H] value of the model was changed at each iteration according to the output of the abundance analysis. The Local Thermodynamic Equilibrium (LTE) program MOOG \citep{1973ApJ...184..839S} was used for the abundance analysis. Final microturbulence and metallicities are reported in Table \ref{Tabla2}.

Another standard technique to get [Fe/H] from the spectra when T$_{\text{eff}}$ and log(g) are known is the spectral synthesis method, which is suitable for low S/N and/or limited wavelength range data. In this case, v$_{\rm t}$ is obtained from a relation that involves log(g) and also T$_{\text{eff}}$ in some cases. We decided to also apply this second method to our spectra as a further check. We used the atmospheric parameters obtained from the HST data and two relations for microturbulence. The first is the empirical relation from \citet{2008A&A...490..625M} that relates log(g) with v$_{\rm t}$ for RGB stars in GCs, and the other is the theoretical relation from \citet{Mott2020} that was obtained from 3D-NLTE models and that relates log(g) and T$_{\text{eff}}$ with v$_{\rm t}$. As the Fe indicator, we used a strong Fe band at 6137 \AA\ that is a blend of many FeI lines and that is visible even in low S/N spectra (see Figure \ref{Fe_6127.png}). Again, atmospheric models were calculated using the ATLAS9 code \citep{1970SAOSR.309.....K} and the LTE program MOOG \citep{1973ApJ...184..839S} was used for the abundance analysis. The results are reported in Table \ref{Tabla3}.

The third method we used to estimate T$_{\text{eff}}$ and log(g) relies on the abundances obtained from EQW of FeI and FeII lines as measured from the spectra. Microturbulence and [Fe/H] were also determined in the process. First of all, atmospheric models were calculated again using the ATLAS9 code \citep{1970SAOSR.309.....K}, assuming our photometric estimations of T$_{\rm eff}$ and log(g) as initial guesses. We assume also v$_{\text{t}}$=1.5 km/s and [Fe/H]=-1.1 as initial estimations. Then T$_{\rm eff}$, log(g), and v$_{\rm t}$ were re-adjusted and new atmospheric models calculated in an interactive way in order to remove trends in excitation potential and reduced EQW versus abundance in order to derive T$_{\rm eff}$ and v$_{\rm t}$, respectively, and to satisfy the ionization equilibrium for log(g). About 100 FeI lines (depending on the S/N of the spectrum) and 4 FeII lines were used for the latter purpose. The [Fe/H] value of the model was changed at each iteration according to the output of the abundance analysis. The LTE program MOOG \citep{1973ApJ...184..839S} was used for this abundance analysis as in the previous methods. Results of this third method are reported in Table \ref{Tabla4}.

\subsection{Comparison of the different techniques} 

\begin{figure}
\centering
\includegraphics[width=3.6in,height=3.0in]{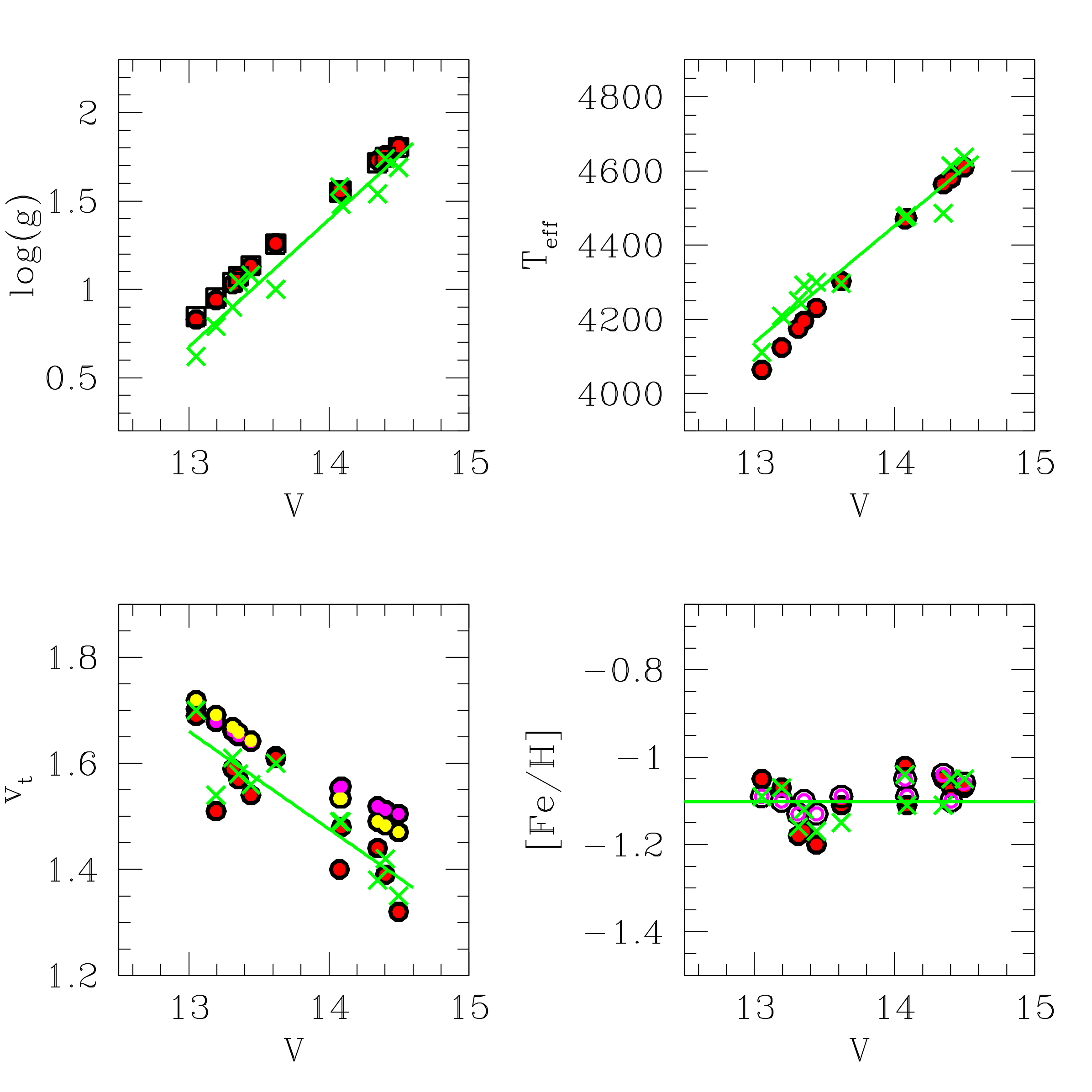}
\caption{Comparison of the estimated atmospheric parameters as a function of V magnitude. The upper panels show log(g) and T$_{\text{eff}}$ obtained from the isochrone-fitting method (red points), from the FeI/FeII equilibrium (green crosses), and log(g) from the canonical equation (black squares). In the lower left panel red points represent v$_{\text{t}}$ obtained from EQW of FeI lines and assuming the isochrone-fitting method parameters, while yellow and magenta points represent microturbulences obtained from the \citet{2008A&A...490..625M} and \citet{Mott2020} relations, respectively. Green crosses show v$_{\text{t}}$ as obtained from EQW of FeI lines and assuming the FeI/FeII equilibrium. In the lower right panel red points represent metallicities obtained using T$_{\text{eff}}$ and log(g) from the isochrone method and v$_{\text{t}}$ and Fe from the EQW of FeI lines. Magenta points represent [Fe/H] obtained using T$_{\text{eff}}$ and log(g) from the  isochrone method and v$_{\text{t}}$ and Fe from the \citet{Mott2020} relation and spectral synthesis method, respectively. Green points represent metallicities obtained using atmospheric parameters from the FeI/FeII balance, and Fe from EQW of FeI lines. The green line is the linear fit to the green crosses.}
\label{Teff_logg.png}
\end{figure}

We first compare the atmospheric parameters we obtained using the different techniques explained in the previous section. This is useful to know and quantify possible discrepancies that might exist between the different methods, and to choose the most appropriate depending on the kind of data one is analyzing. We start by comparing T$_{\text{eff}}$ and log(g). For this purpose, in the upper panels of Figure \ref{Teff_logg.png} we report log(g) vs. V magnitude (left panel) and T$_{\text{eff}}$ vs. V magnitude (right panel) for our targets. V magnitude has an error that is negligible in the scale of the figure, and any discrepancy visible between the points is due only to discrepancies in gravity and temperature respectively. In the upper panels, red points represent gravities and temperatures obtained using the isochrone-fitting method, green crosses are the same parameters obtained using the FeI and FeII equilibrium, while the black squares on the upper left panel represent log(g) obtained from the canonical equation. The green line is the linear fit to the green crosses. We see that the best agreement between CMD fitting and FeI/II balance happens for the fainter and hotter targets. For bright stars, the FeI/II lines balance method gives gravities and temperatures that are slightly lower and higher (respectively) compared with the CMD fitting method. The discrepancy is of order -0.1 dex in log(g) and +40 K in T$_{\text{eff}}$. On the other hand, the gravity obtained from the canonical equation agrees very well with those obtained by isochrone fitting.

For microturbulence, the lower left panel shows v$_{\rm t}$ as obtained from EQW of FeI lines and assuming the isochrone method parameters (red points), from the empirical \citet{2008A&A...490..625M} relation (yellow points), from the theoretical \citet{Mott2020} relation (magenta points), and from EQW of FeI lines and assuming the FeI/II lines balance method parameters (green crosses). We underline the very good agreement between the empirical and the theoretical relation in spite of the fact that they were obtained in two completely different ways. Since the relation from \citet{Mott2020} depends also on the temperature, this means that it can be safely used to obtained the microturbulence in evolutionary stages different from the RGB (i.e. AGB, RC, SGB, MS), where the relation from \citet{2008A&A...490..625M} is not valid. On the other hand, v$_{\rm t}$ obtained from EQW of FeI and parameters from the isochrone-fitting method agrees well with that obtained from EQW of FeI and parameters from the FeI/II lines balance, but are slightly lower (about -0.1 km/s) than that obtained from the \citet{Mott2020} relation.

The lower right panel shows finally the impact of the different methods used to obtain the parameters and to estimate the abundance from iron lines on the final [Fe/H] measurement. Here red points represent the metallicities obtained using T$_{\text{eff}}$ and log(g) from the isochrone method and microturbulence and metallicity from the EQW of FeI lines. Magenta points represent the metallicities obtained using T$_{\text{eff}}$ and log(g) from the isochrone method but the microturbulence from the \citet{Mott2020} relation and the metallicity from the fit of the 6137 \AA\ iron band. Green points represent the metallicities obtained using T$_{\text{eff}}$, log(g), and v$_{\rm t}$ from the FeI/II lines balance, and the metallicity from the EQW of FeI lines. We see that, in spite of the fact that the three methods used to get the final [Fe/H] values are partially independent from each other, the metallicities agree very well and the systematic difference is very low. The first and third methods give [Fe/H]$_1$=-1.10$\pm0.02$ and [Fe/H]$_3$=-1.10$\pm0.01$ respectively, while the second returns [Fe/H]$_2$=-1.09$\pm0.01$. The only real difference is in the dispersion. The second method returns the lowest spread, with $\sigma_{\text{[Fe/H]}_2}=0.03\pm0.01$, while the first and third give $\sigma_{\text{[Fe/H]}_1}=0.06\pm0.01$ and $\sigma_{\text{[Fe/H]}_3}=0.05\pm0.01$ respectively.

Our conclusion is that, in spite of the fact that the atmospheric parameters obtained from the different methods of analysis we applied show a certain degree of disagreement (however below 50 K in temperature, 0.1 dex in gravity and 0.1 km/s in microturbulence), the final [Fe/H] appears not to be affected by any systematics, at least differentially. We underline also the fact that the quickest method we applied, i.e. T$_{\text{eff}}$ and log(g) from the isochrone-fitting method, microturbulence from the \citet{Mott2020} relation and the metallicity from the fit of the 6137 \AA\ iron band, is also the one that returns the data with the lowest spread.

\section{RESULTS and COMPARISON with LITERATURE} \label{Results}

The mean iron content values we obtained with the three methods are:
$$ [Fe/H]_{1} = -1.10 \pm 0.02 $$
$$ [Fe/H]_{2} = -1.09 \pm 0.01 $$
$$ [Fe/H]_{3} = -1.10 \pm 0.01 $$

with a dispersion of:
$$ \sigma_{[Fe/H]_{1}} = 0.06 \pm 0.01 $$
$$ \sigma_{[Fe/H]_{2}} = 0.03 \pm 0.01 $$
$$ \sigma_{[Fe/H]_{3}} = 0.05 \pm 0.01 $$

The typical internal error on temperature determination is between 20 and 40 K, depending on the method, while the errors on gravity and microturbulence are below 0.1 dex and 0.05 km/s respectively. If we apply the same procedure described in \citet{Villanova2013} and \citet{2008A&A...490..625M} for the error calculation we obtain $\sigma_{\text{TOT}}=0.05$ for our estimate of the total internal error in metallicity.
Comparing this value with the observed spreads, we see that there is no room for an intrinsic metallicity variation in NGC 362. Even more, the second method delivers an iron spread that is 1.7 - 2 times lower that the other two, strengthening our statement.

Figure \ref{ChromosomeMap1.png} shows the chromosome map of NGC 362 with the typical separation between 1G (lower) and 2G (upper) stars divided by the magenta dashed line. The significant distribution of red (anomalous) stars located on the right part of the main loci of 1G and 2G (normal) stars is the cause for the classification of NGC 362 as Type II \citep{Milone17}. According to their scenario, the anomalous stars could have an iron content higher than the normal population. As we mentioned earlier, our target stars were chosen to cover a wide range in the chromosome map in order to  allow us to check if an iron spread is indeed present. We carefully selected an equal number of normal and anomalous stars in order to best test the possibility of a predicted metallicity variation between them. In the end, we obtained very good spectra for 6 normal and 5 anomalous members of NGC 362. A close inspection of the chromosome map, reveals that the fraction of red stars classified as anomalous represent the 5.5\% with respect to the total number of stars. The derived fraction is in good agreement with the $\sim$7.5\% of stars possibly having enhanced metal abundances in this cluster according to the \citet{Milone17} scenario.

In spite of the classification of NGC 362 as a Type II cluster, the measured iron dispersion does not support the existence of a significant iron spread in our sample. In our sample we have 5 anomalous stars, located on the red side of the main 1G and 2G populations and for this reason, according to \citet{Milone17}, they could show a [Fe/H] value significantly larger (0.1-0.2 dex) than the normal stars. However, their spectroscopically-derived [Fe/H] content spans the range -1.04 to -1.15 dex, insignificantly different from the spread of the 6 normal stars (-1.05 to -1.17 dex). What's more, looking in detail at the spectroscopic results of our target stars in each group, we find a mean metallicity of [Fe/H]$_n$=-1.10$\pm$0.02 with a spread of $\sigma_n$=0.04$\pm$0.01 for the 6 normal stars vs. [Fe/H]$_a$=-1.10$\pm$0.02 with a spread of $\sigma_a$=0.05$\pm$0.01 for the 5 anomalous stars. This confirms our main result: that NGC 362 does not host stars with an intrinsic iron spread, and in particular there is no metallicity difference between normal and anomalous stars.

We now discuss how our result compares to results in the literature. It is worth nothing that our study is the first one that purposefully compares metallicities in normal vs. anomalous stars. \citet{Worley10} analyzed the high-resolution spectra of 13 giant stars in NGC 362 taken with the UVES spectrograph. Fe, light (Sr, Y, Zr) and heavy (Ba, Nd, La) s-process element abundances were derived from EQWs. They found an average metallicity of [Fe/H]=-1.21$\pm$0.09 and a comparable abundance enhancement in the light and heavy elements for the sample with values of [ls/Fe]=0.32$\pm$0.10 dex and [hs/Fe]=0.46$\pm$0.09 dex, respectively. The small spread in these values and the lack of trend of [hs/ls] with metallicity, indicates a clear homogeneity in the s-process elemental abundance enhancements for NGC 362. It also indicates that the spread in the overall derived metallicity for these stars is due to the analysis rather than intrinsic differences between the stars. 

\citet{Carretta13} enlarged the sample of stars analyzed in NGC 362 with moderate-high resolution FLAMES GIRAFFE+UVES spectroscopy for both first and second generation RGB stars, and used it to derive abundances of proton-capture, $\alpha$-capture, Fe-peak and neutron-capture elements from EQWs. The mean metallicity found with UVES and GIRAFFE spectra was [Fe/H]=-1.17$\pm$0.05 dex (for a sample of 14 stars) and [Fe/H]=-1.17$\pm$0.04 dex (for 90 targets), respectively. The authors reported no evidence for an internal Fe spread and concluded that NGC 362 behaves like an ordinary cluster in terms of elements involved at high temperatures in the H-burning process. Nevertheless they discovered the presence of an additional, poorly populated (e.g. accounting for only $\sim$6\% of the total cluster population) red RGB sequence which appears to be enriched in s-process elements, particularly in Ba.

\citet{Bailin19} published an up-to-date homogeneous catalog of intrinsic Fe spreads of RGB stars in 55 Milky Way GCs, for which enough precise spectroscopic observations of iron lines were available. Bailin reanalyzed the spectroscopic data from \citet{Worley10} for NGC 362 and measured an average metallicity of [Fe/H]=-1.21$\pm$0.02 with a dispersion of $\sigma$=0.07. He did not find any trace of an intrinsic iron spread, but reported a trend of larger metallicity dispersion with higher luminosity for GCs with $L>10^5 L_{\odot}$. 

\citet{Marino2019} explored the chemical properties of MSPs in 29 GCs using the photometric data base from the HST UV Legacy Survey supplemented by ground based photometry, and coupled it to spectroscopic chemical abundances of the same stars from the literature. The authors combined the elemental abundances of Li, N, O, Na, Mg, Al, Si, K, Ca, Fe, and Ba and the chromosome maps of the GCs to characterize the chemical composition of the distinct stellar populations and calculate the average abundance of 1G and 2G stars for each analysed element. Regarding to NGC 362, Marino et al. found an iron abundance of [Fe/H]=-1.18$\pm$0.04 and [Fe/H]=-1.17$\pm$0.05 for a sample of 6 normal and 3 anomalous stars, respectively, which is in good agreement with our results. They also reported that NGC 362 apparently does not follow the [Ba/Fe] vs. [Fe/H] trend observed in other Type II GCs.

More recently, \citet{Meszaros20} investigated the Fe, C, N, O, Mg, Al, Si, K, Ca, Ce and Nd abundances in 31 GCs from high-resolution spectra observed by the SDSS-IV APOGEE-2 survey. They collected spectra for 56 RGB stars in NGC 362 and used the BACCHUS code \citep{Masseron16} to determine metallicities and abundances. The mean iron content obtained was [Fe/H]=-1.03$\pm$0.06 with a dispersion of $\sigma$=0.08. Mészáros et al. found no indication for an internal metallicity variation, however they have not investigated if any Fe spread exists between normal and anomalous stars. When looking at other elements, the authors found clear Al-O anticorrelation and Al-N correlation with a weakly present Al-Mg anticorrelation, in addition to a small Al spread of [Al/Fe]$_{\text{scatter}}$=0.24. On the other hand, a uniform and constant C+N+O content was measured, as well as a clear s-process element enhancement, confirming the study of \citet{Carretta13}. A future paper will report on abundances of a large number of species, including s-process elements.

Thus, there is an array of existing studies of the metallicities of a significant number of stars based on high-resolution spectroscopy. Although our's is the first to explicitly select purported normal and anomalous stars a priori, other studies undoubtedly included both types of stars in their large samples, especially the studies of \citet{Carretta13} and \citet{Meszaros20}. The concatenation of this observational evidence does not support one of the hypothesis of \citet{Milone17} that NGC 362 is characterised by an intrinsic iron spread, with our study being the most decisive, at least in terms of preselecting normal and anomalous stars. It is extremely unlikely that a real spread was simply missed due to improper and/or small samples or large errors.

\citet{Lardo16} reported that spurious spreads observed in the s-process rich group of stars are derived in anomalous GCs (i.e. clusters with spreads in s-process elements and possibly C+N+O, double SGBs and split RGBs) when spectroscopic gravities are adopted. Such clusters would be indeed mono-metallic when FeII lines and photometric gravities were used in the analysis. \citet{Carretta13} adopted photometric stellar gravities and they did not report an [Fe/H] spread for stars in NGC 362, which is consistent with the hypothesis that spectroscopic gravities might yield artificial iron spreads. However, note that \citet{Munoz21} found an intrinsic spread in the Type II GC NGC 1261, using both photometric and spectroscopic gravities.

Given the importance of understanding Type II GCs formation and evolution, it is of great interest to investigate with high-resolution spectra other Type II GCs with presumed metallicity spreads. Of the ten Type II GCs identified by \citet{Milone17}, \citet{Meszaros20} derived abundances and discussed intrinsic iron spreads for NGC 362, NGC 1851, $\omega$ Centauri, NGC 6388, M22 and M2, while \citet{Marino2019} studied NGC 5286 and M54. \citet{Marino18,Marino2021} analyzed the lowest mass Type II GC NGC 6934 and more recently \citet{Munoz21} performed a detailed spectroscopic metallicity analysis of the second lowest mass GC NGC 1261. From the existing data it is not clear that all of these GCs presumed to have Fe spreads may in fact possess them. \citet{Meszaros20} do not confirm the metallicity spreads previously observed in NGC 1851, M22 and M2 \citep{Carretta10,Marino,Marino2019} and neither detected variations in the [Fe/H] abundances of NGC 362 and NGC 6388. The only real spreads found were in $\omega$ Centauri and M54 \citep{Carretta2010b,Marino2019}. However, both of these objects are classified as the remnant cores of tidally disrupted (or disrupting) dwarf galaxies and hence have a different chemical evolution history from typical GCs. Looking at the spectroscopic results from the catalog of \citet{Bailin19}, we see that Fe variations were confirmed in M54, $\omega$ Centauri and NGC 5286 clusters. Moreover, he reported that the former has the second largest Fe spread of his study, with the largest belonging to $\omega$ Centauri. On the contrary, the Type II GC M15, classified by \citet{Nardiello2018}, does not present evidence of an iron abundance spread. Finally, regarding the two lowest mass Type II GCs, NGC 1261 and NGC 6934, they do both show intrinsic metallicity variations \citep{Munoz21,Marino18}, although both studies used a small sample of stars. 

The existence of an intrinsic real Fe spread in M22 has been highly debated in the literature since spectroscopic and photometric studies based on small samples of stars have yielded conflicting results. \citet{Cohen81} and \citet{Gratton82} conclude that no significant metallicity variation is present in this Type II GC, while \citet{Pilachowski82} and \citet{Lehnert91} found evidence of an intrinsic Fe spread, with -1.4<[Fe/H]<-1.9. \citet{Marino2009} confirmed the existence of an intrinsic Fe variation from the analysis of high-resolution UVES spectra of 17 stars, reporting that the difference between the mean Fe abundances of the two identified groups of stars is 0.14 dex. More recently, \citet{Marino} analyzed high-resolution spectra of 35 giant stars, finding that M22 hosts at least two groups of stars characterized by substantial star-to-star metallicity scatter  (-2.0$\lesssim$[Fe/H]$\gtrsim$-1.6), s-process and CNO-cycle elements. Thus, this cluster was able to retain the typical high-mass AGB ejecta, as well as the SNe ejecta. \citet{Mucciarelli2015} presented a new high-resolution spectra analysis of 17 giant stars that \citet{Marino2009} used to provide evidence in support to an intrinsic metallicity spread in M22, and found that when surface gravities are derived spectroscopically the [Fe/H] distribution spans $\sim$0.5 dex, but when photometric gravities are adopted, the [FeII/H] distribution shows no evidence of an iron spread, concluding that M22 is a normal GC and ruling out that it has retained the supernovae ejecta in its gravitational potential well. The later spectroscopic study of \citet{Meszaros20} does not observe a significant Fe variation in M22 either, reporting a [Fe/H]=-1.52$\pm$0.02 with a dispersion of $\sigma$=0.11, expected from the internal errors, based on a sample of 20 high S/N spectra observed with APOGEE.

The analysis of \citet{Munoz21} does not find any relation between intrinsic metallicity spread and either GC mass, the fraction of Type II stars (N$_{\text{TypeII}}$/N$_{\text{Tot}}$) or the fraction of 1G stars (N$_1$/N$_{\text{Tot}}$). In particular, NGC 362 does not stand out in any of these plots. Its mass is greater than that of either of the two very low mass Type II GCs, NGC 1261 and NGC 6934, both of which appear to have real spreads, comparable to that of intermediate mass Type IIs that do not appear to have spreads, and of course much less than the masses of NGC 6715 and $\omega$ Centauri.

\section{Summary and Conclusions} \label{Summary and conclusions}

In this paper we have presented a chemical characterization of the GC NGC 362, which is classified as a Type II GC by \citet{Milone17}, focusing on its iron content. \citet{Marino18} suggests that Type II GCs possess an intrinsic variation in metallicity, with anomalous stars (those lying both on the reddest RGB as well as to the red of the chromosome map) predicted to have higher metallicity and higher iron than normal main RGB or chromosome map stars. We observed twelve RGB stars using the high-resolution MIKE spectrograph mounted at the Magellan-Clay telescope. The stars were carefully selected from HST UV Legacy data to include a comparable sample of normal and anomalous stars. One of the anomalous stars proved to be a less-likely proper motion member, as well as overly spatially crowded, and was discarded. From the spectra and HST and GAIA photometry, we determined the atmospheric parameters and measured the iron content of all of our sample with three different methods, and then derived the mean metallicity and its dispersion for each method:

\begin{itemize}
    \item First method: The iron content was obtained using T$_{\text{eff}}$ and log(g) from the isochrone-fitting method, and v$_\text{t}$ and metallicity from the EQW of FeI lines. This method returns [Fe/H]$_1=-1.10\pm 0.02$ and $\sigma_{\text{[Fe/H]}_1}=0.06\pm 0.01$.
    
    \item Second method: [Fe/H] was determined using T$_{\text{eff}}$ and log(g) from the CMD isochrones fitting method and v$_\text{t}$ from the \citet{Mott2020} relation. We used spectral synthesis to obtain the metallicity from the fit of the 6137 \AA\ iron band. This method returns [Fe/H]$_2=-1.09\pm0.01$ and $\sigma_{\text{[Fe/H]}_2}=0.03\pm0.01$. 
    
    \item Third method: The iron abundance was obtained using T$_{\text{eff}}$, log(g) and v$_\text{t}$ from the FeI and FeII lines balance, and the metallicity from the EQW of FeI lines. This method returns [Fe/H]$_3=-1.10\pm0.01$ and $\sigma_{\text{[Fe/H]}_3}=0.05\pm0.01$.
    
    \item The error analysis gives an internal error of 0.05 dex. Comparing these values, we conclude that NGC 362 does not show any trace of an internal iron spread. 
    
    \item The three methods we used return the same mean iron abundance within the errors, in spite of the fact that they are partially independent from each other. In particular, we find no significant difference between the metallicity of the normal vs. anomalous stars, therefore we do not confirm one of the hypothesis of \citet{Milone17}. Additionally, previous high-resolution spectroscopic studies of large samples of stars in NGC 362 in the literature did not report any evidence for an intrinsic metallicity variation, endorsing our results.
    
    \item The recent paper by \citet{Munoz21} does not find any significant relation between intrinsic metallicity spread and either GC mass, (N$_{\text{TypeII}}$/N$_{\text{Tot}}$) or (N$_1$/N$_{\text{Tot}}$) in Type II GCs. Indeed, it suggests that a number of well studied Type II GCs in fact do not possess intrinsic Fe abundance spreads, and therefore the scenario of \citet{Milone17} is not fully borne out. Although several GCs with classically-known metallicity spreads, viz. $\omega$ Centauri and NGC 6715, are also Type II GCs, and several low mass Type II GCs also appear to have intrinsic metallicity spreads (NGC 1261 and NGC 6934), the other Type II GCs do not clearly demonstrate this phenomenon. 
\end{itemize}

\section{Data Availability}

The data underlying this article will be shared on reasonable request to the corresponding author.

\section{Acknowledgements}

S.V. gratefully acknowledges the support provided by FONDECYT Regular No. 1220264. S.V. also gratefully acknowledges support by the Agencia Nacional de Investigación y Desarrollo (ANID) BASAL projects ACE210002 and FB210003. D.G. gratefully acknowledges support from the Chilean Centro de Excelencia en Astrof\'isica y Tecnolog\'ias Afines (CATA) BASAL grant AFB-170002. D.G. also acknowledges financial support from the Direcci\'on de Investigaci\'on y Desarrollo de la Universidad de La Serena through the Programa de Incentivo a la Investigaci\'on de Acad\'emicos (PIA-DIDULS).

\bibliographystyle{mnras}
\bibliography{bibliography.bib}

\begin{thebibliography}{}
\makeatletter
\relax
\def\mn@urlcharsother{\let\do\@makeother \do\$\do\&\do\#\do\^\do\_\do\%\do\~}
\def\mn@doi{\begingroup\mn@urlcharsother \@ifnextchar [ {\mn@doi@}
  {\mn@doi@[]}}
\def\mn@doi@[#1]#2{\def\@tempa{#1}\ifx\@tempa\@empty \href
  {http://dx.doi.org/#2} {doi:#2}\else \href {http://dx.doi.org/#2} {#1}\fi
  \endgroup}
\def\mn@eprint#1#2{\mn@eprint@#1:#2::\@nil}
\def\mn@eprint@arXiv#1{\href {http://arxiv.org/abs/#1} {{\tt arXiv:#1}}}
\def\mn@eprint@dblp#1{\href {http://dblp.uni-trier.de/rec/bibtex/#1.xml}
  {dblp:#1}}
\def\mn@eprint@#1:#2:#3:#4\@nil{\def\@tempa {#1}\def\@tempb {#2}\def\@tempc
  {#3}\ifx \@tempc \@empty \let \@tempc \@tempb \let \@tempb \@tempa \fi \ifx
  \@tempb \@empty \def\@tempb {arXiv}\fi \@ifundefined
  {mn@eprint@\@tempb}{\@tempb:\@tempc}{\expandafter \expandafter \csname
  mn@eprint@\@tempb\endcsname \expandafter{\@tempc}}}

\bibitem[\protect\citeauthoryear{{Alonso}, {Arribas}  \&
  {Mart{\'\i}nez-Roger}}{{Alonso} et~al.}{1999}]{1999A&AS..140..261A}
{Alonso} A.,  {Arribas} S.,   {Mart{\'\i}nez-Roger} C.,  1999, \mn@doi [\aaps]
  {10.1051/aas:1999521}, \href
  {https://ui.adsabs.harvard.edu/abs/1999A&AS..140..261A} {140, 261}

\bibitem[\protect\citeauthoryear{Bailin}{Bailin}{2019}]{Bailin19}
Bailin J.,  2019, \mn@doi [The Astrophysical Journal Supplement Series]
  {10.3847/1538-4365/ab4812}, 245, 5

\bibitem[\protect\citeauthoryear{{Bastian} \& {Lardo}}{{Bastian} \&
  {Lardo}}{2018}]{Bastian2018}
{Bastian} N.,  {Lardo} C.,  2018, \mn@doi [\araa]
  {10.1146/annurev-astro-081817-051839}, \href
  {https://ui.adsabs.harvard.edu/abs/2018ARA&A..56...83B} {56, 83}

\bibitem[\protect\citeauthoryear{Bastian, Lamers, de Mink, Longmore, Goodwin
  \& Gieles}{Bastian et~al.}{2013}]{Batian2013}
Bastian N.,  Lamers H. J. G. L.~M.,  de Mink S.~E.,  Longmore S.~N.,  Goodwin
  S.~P.,   Gieles M.,  2013, \mn@doi [Monthly Notices of the Royal Astronomical
  Society] {10.1093/mnras/stt1745}, 436, 2398

\bibitem[\protect\citeauthoryear{{Baumgardt} \& {Hilker}}{{Baumgardt} \&
  {Hilker}}{2018}]{2018MNRAS.478.1520B}
{Baumgardt} H.,  {Hilker} M.,  2018, \mn@doi [\mnras] {10.1093/mnras/sty1057},
  \href {https://ui.adsabs.harvard.edu/abs/2018MNRAS.478.1520B} {478, 1520}

\bibitem[\protect\citeauthoryear{Baumgardt, Hilker, Sollima  \&
  Bellini}{Baumgardt et~al.}{2019}]{Baumgardt19}
Baumgardt H.,  Hilker M.,  Sollima A.,   Bellini A.,  2019, \mn@doi [Monthly
  Notices of the Royal Astronomical Society] {10.1093/mnras/sty2997}, 482, 5138

\bibitem[\protect\citeauthoryear{Bernstein, Shectman, Gunnels, Mochnacki  \&
  Atheya}{Bernstein et~al.}{2003}]{Bernstein}
Bernstein R.,  Shectman S.,  Gunnels S.,  Mochnacki S.,   Atheya A.,  2003,
  \mn@doi [Proceedings of SPIE - The International Society for Optical
  Engineering] {10.1117/12.461502}, 4841

\bibitem[\protect\citeauthoryear{{Bressan}, {Marigo}, {Girardi}, {Salasnich},
  {Dal Cero}, {Rubele}  \& {Nanni}}{{Bressan}
  et~al.}{2012}]{2012MNRAS.427..127B}
{Bressan} A.,  {Marigo} P.,  {Girardi} L.,  {Salasnich} B.,  {Dal Cero} C.,
  {Rubele} S.,   {Nanni} A.,  2012, \mn@doi [\mnras]
  {10.1111/j.1365-2966.2012.21948.x}, \href
  {https://ui.adsabs.harvard.edu/abs/2012MNRAS.427..127B} {427, 127}

\bibitem[\protect\citeauthoryear{{Carretta} \& {Bragaglia}}{{Carretta} \&
  {Bragaglia}}{2021}]{Carretta2021}
{Carretta} E.,  {Bragaglia} A.,  2021, \mn@doi [\aap]
  {10.1051/0004-6361/202142563}, \href
  {https://ui.adsabs.harvard.edu/abs/2022A&A...659A.122C} {659, A122}

\bibitem[\protect\citeauthoryear{{Carretta} et~al.,}{{Carretta}
  et~al.}{2009a}]{Carretta}
{Carretta} et~al., 2009a, \mn@doi [A\&A] {10.1051/0004-6361/200912096}, 505,
  117

\bibitem[\protect\citeauthoryear{{Carretta}, {Bragaglia, A.}, {Gratton, R.}  \&
  {Lucatello, S.}}{{Carretta} et~al.}{2009b}]{Carretta2009a}
{Carretta} {Bragaglia, A.} {Gratton, R.}  {Lucatello, S.} 2009b, \mn@doi [A\&A]
  {10.1051/0004-6361/200912097}, 505, 139

\bibitem[\protect\citeauthoryear{{Carretta}, {Bragaglia, A.}, {Gratton, R.},
  {D'Orazi, V.}  \& {Lucatello, S.}}{{Carretta} et~al.}{2009c}]{Carretta2009}
{Carretta} {Bragaglia, A.} {Gratton, R.} {D'Orazi, V.}  {Lucatello, S.} 2009c,
  \mn@doi [A\&A] {10.1051/0004-6361/200913003}, 508, 695

\bibitem[\protect\citeauthoryear{{Carretta} et~al.,}{{Carretta}
  et~al.}{2010a}]{Carretta2010b}
{Carretta} et~al., 2010a, \mn@doi [A\&A] {10.1051/0004-6361/201014924}, 520,
  A95

\bibitem[\protect\citeauthoryear{{Carretta} et~al.,}{{Carretta}
  et~al.}{2010b}]{Carretta10}
{Carretta} E.,  et~al., 2010b, \mn@doi [\apjl] {10.1088/2041-8205/722/1/L1},
  \href {https://ui.adsabs.harvard.edu/abs/2010ApJ...722L...1C} {722, L1}

\bibitem[\protect\citeauthoryear{{Carretta}, {Lucatello, S.}, {Gratton, R. G.},
  {Bragaglia, A.}  \& {D\'{}Orazi, V.}}{{Carretta} et~al.}{2011}]{Carretta2011}
{Carretta} {Lucatello, S.} {Gratton, R. G.} {Bragaglia, A.}  {D\'{}Orazi, V.}
  2011, \mn@doi [A\&A] {10.1051/0004-6361/201117269}, 533, A69

\bibitem[\protect\citeauthoryear{{Carretta} et~al.,}{{Carretta}
  et~al.}{2013}]{Carretta13}
{Carretta} et~al., 2013, \mn@doi [A\&A] {10.1051/0004-6361/201321905}, 557,
  A138

\bibitem[\protect\citeauthoryear{{Charbonnel}}{{Charbonnel}}{2016}]{Charbonnel2016}
{Charbonnel} C.,  2016, in EAS Publications Series. pp 177--226 (\mn@eprint
  {arXiv} {1611.08855}), \mn@doi{10.1051/eas/1680006}

\bibitem[\protect\citeauthoryear{{Cohen}}{{Cohen}}{1981}]{Cohen81}
{Cohen} J.~G.,  1981, \mn@doi [\apj] {10.1086/159097}, \href
  {https://ui.adsabs.harvard.edu/abs/1981ApJ...247..869C} {247, 869}

\bibitem[\protect\citeauthoryear{Collaboration, Brown, Vallenari, Prusti, de
  Bruijne, Babusiaux  \& Biermann}{Collaboration et~al.}{2020}]{GAIA2020}
Collaboration G.,  Brown A. G.~A.,  Vallenari A.,  Prusti T.,  de Bruijne J.
  H.~J.,  Babusiaux C.,   Biermann M.,  2020, Gaia Early Data Release 3:
  Summary of the contents and survey properties (\mn@eprint {arXiv}
  {2012.01533})

\bibitem[\protect\citeauthoryear{{D'Antona}, {Vesperini}, {D'Ercole},
  {Ventura}, {Milone}, {Marino}  \& {Tailo}}{{D'Antona}
  et~al.}{2016}]{D'Antona2016}
{D'Antona} F.,  {Vesperini} E.,  {D'Ercole} A.,  {Ventura} P.,  {Milone} A.~P.,
   {Marino} A.~F.,   {Tailo} M.,  2016, \mn@doi [\mnras]
  {10.1093/mnras/stw387}, \href
  {https://ui.adsabs.harvard.edu/abs/2016MNRAS.458.2122D} {458, 2122}

\bibitem[\protect\citeauthoryear{D'Ercole, Vesperini, D'Antona, McMillan  \&
  Recchi}{D'Ercole et~al.}{2008}]{DErcole}
D'Ercole A.,  Vesperini E.,  D'Antona F.,  McMillan S. L.~W.,   Recchi S.,
  2008, \mn@doi [Monthly Notices of the Royal Astronomical Society]
  {10.1111/j.1365-2966.2008.13915.x}, 391, 825

\bibitem[\protect\citeauthoryear{{Decressin}, {Meynet, G.}, {Charbonnel, C.},
  {Prantzos, N.}  \& {Ekstr\"om, S.}}{{Decressin} et~al.}{2007}]{Decressin}
{Decressin} {Meynet, G.} {Charbonnel, C.} {Prantzos, N.}  {Ekstr\"om, S.} 2007,
  \mn@doi [A\&A] {10.1051/0004-6361:20066013}, 464, 1029

\bibitem[\protect\citeauthoryear{Denissenkov \& Hartwick}{Denissenkov \&
  Hartwick}{2014}]{Denissenkov2014}
Denissenkov P.~A.,  Hartwick F. D.~A.,  2014, \mn@doi [Monthly Notices of the
  Royal Astronomical Society: Letters] {10.1093/mnrasl/slt133}, 437, L21

\bibitem[\protect\citeauthoryear{Ferraro et~al.,}{Ferraro
  et~al.}{2009}]{Ferraro2009}
Ferraro F.~R.,  et~al., 2009, \mn@doi [Nature] {10.1038/nature08581}, 462,
  483–486

\bibitem[\protect\citeauthoryear{{Gratton}}{{Gratton}}{1982}]{Gratton82}
{Gratton} R.~G.,  1982, \aap, \href
  {https://ui.adsabs.harvard.edu/abs/1982A&A...115..171G} {115, 171}

\bibitem[\protect\citeauthoryear{{Gratton}, {Johnson, C. I.}, {Lucatello, S.},
  {D\'{}Orazi, V.}  \& {Pilachowski, C.}}{{Gratton} et~al.}{2011}]{Gratton2011}
{Gratton} {Johnson, C. I.} {Lucatello, S.} {D\'{}Orazi, V.}  {Pilachowski, C.}
  2011, \mn@doi [A\&A] {10.1051/0004-6361/201117093}, 534, A72

\bibitem[\protect\citeauthoryear{{Gratton}, {Carretta}  \&
  {Bragaglia}}{{Gratton} et~al.}{2012}]{Gratton2012}
{Gratton} R.~G.,  {Carretta} E.,   {Bragaglia} A.,  2012, \mn@doi [\aapr]
  {10.1007/s00159-012-0050-3}, \href
  {https://ui.adsabs.harvard.edu/abs/2012A&ARv..20...50G} {20, 50}

\bibitem[\protect\citeauthoryear{{Gratton}, {Bragaglia}, {Carretta}, {D'Orazi},
  {Lucatello}  \& {Sollima}}{{Gratton} et~al.}{2019}]{Gratton2019}
{Gratton} R.,  {Bragaglia} A.,  {Carretta} E.,  {D'Orazi} V.,  {Lucatello} S.,
   {Sollima} A.,  2019, \mn@doi [\aapr] {10.1007/s00159-019-0119-3}, \href
  {https://ui.adsabs.harvard.edu/abs/2019A&ARv..27....8G} {27, 8}

\bibitem[\protect\citeauthoryear{{Harris}}{{Harris}}{1996}]{1996AJ....112.1487H}
{Harris} W.~E.,  1996, \mn@doi [\aj] {10.1086/118116}, \href
  {https://ui.adsabs.harvard.edu/abs/1996AJ....112.1487H} {112, 1487}

\bibitem[\protect\citeauthoryear{Harris}{Harris}{2010}]{harris2010new}
Harris W.~E.,  2010, A New Catalog of Globular Clusters in the Milky Way
  (\mn@eprint {arXiv} {1012.3224})

\bibitem[\protect\citeauthoryear{Harris}{Harris}{2018}]{Harris2018TransformationOH}
Harris W.~E.,  2018, The Astronomical Journal, 156, 296

\bibitem[\protect\citeauthoryear{{Johnson} \& {Pilachowski}}{{Johnson} \&
  {Pilachowski}}{2010}]{Johnson2010}
{Johnson} C.~I.,  {Pilachowski} C.~A.,  2010, \mn@doi [\apj]
  {10.1088/0004-637X/722/2/1373}, \href
  {https://ui.adsabs.harvard.edu/abs/2010ApJ...722.1373J} {722, 1373}

\bibitem[\protect\citeauthoryear{Kelson}{Kelson}{2003}]{Kelson_2003}
Kelson D.,  2003, \mn@doi [Publications of the Astronomical Society of the
  Pacific] {10.1086/375502}, 115, 688–699

\bibitem[\protect\citeauthoryear{Kelson, Illingworth, Dokkum  \& Franx}{Kelson
  et~al.}{2000}]{Kelson_2000}
Kelson D.,  Illingworth G.,  Dokkum P.~V.,   Franx M.,  2000, The Astrophysical
  Journal, 531, 184

\bibitem[\protect\citeauthoryear{{Kurucz}}{{Kurucz}}{1970}]{1970SAOSR.309.....K}
{Kurucz} R.~L.,  1970, SAO Special Report, \href
  {https://ui.adsabs.harvard.edu/abs/1970SAOSR.309.....K} {309}

\bibitem[\protect\citeauthoryear{Lardo, Mucciarelli  \& Bastian}{Lardo
  et~al.}{2016}]{Lardo16}
Lardo C.,  Mucciarelli A.,   Bastian N.,  2016, \mn@doi [Monthly Notices of the
  Royal Astronomical Society] {10.1093/mnras/stv2802}, 457, 51–63

\bibitem[\protect\citeauthoryear{Lehnert, Bell  \& Cohen}{Lehnert
  et~al.}{1991}]{Lehnert91}
Lehnert M.~D.,  Bell R.~A.,   Cohen J.~G.,  1991, \mn@doi [Astrophysical
  Journal; (USA)] {10.1086/169648}, 367

\bibitem[\protect\citeauthoryear{{Majewski}, {Patterson}, {Dinescu}, {Johnson},
  {Ostheimer}, {Kunkel}  \& {Palma}}{{Majewski} et~al.}{2000}]{Majewski}
{Majewski} S.~R.,  {Patterson} R.~J.,  {Dinescu} D.~I.,  {Johnson} W.~Y.,
  {Ostheimer} J.~C.,  {Kunkel} W.~E.,   {Palma} C.,  2000, in {Noels} A.,
  {Magain} P.,  {Caro} D.,  {Jehin} E.,  {Parmentier} G.,   {Thoul} A.~A.,
  eds,  Liege International Astrophysical Colloquia Vol. 35, Liege
  International Astrophysical Colloquia. p.~619 (\mn@eprint {arXiv}
  {astro-ph/9910278})

\bibitem[\protect\citeauthoryear{{Mar{\'\i}n-Franch}
  et~al.,}{{Mar{\'\i}n-Franch} et~al.}{2009}]{Mar09}
{Mar{\'\i}n-Franch} A.,  et~al., 2009, \mn@doi [\apj]
  {10.1088/0004-637X/694/2/1498}, \href
  {https://ui.adsabs.harvard.edu/abs/2009ApJ...694.1498M} {694, 1498}

\bibitem[\protect\citeauthoryear{{Marino}, {Villanova}, {Piotto}, {Milone},
  {Momany}, {Bedin}  \& {Medling}}{{Marino} et~al.}{2008}]{2008A&A...490..625M}
{Marino} A.~F.,  {Villanova} S.,  {Piotto} G.,  {Milone} A.~P.,  {Momany} Y.,
  {Bedin} L.~R.,   {Medling} A.~M.,  2008, \mn@doi [\aap]
  {10.1051/0004-6361:200810389}, \href
  {https://ui.adsabs.harvard.edu/abs/2008A&A...490..625M} {490, 625}

\bibitem[\protect\citeauthoryear{{Marino}, {Milone}, {Piotto}, {Villanova},
  {Bedin}, {Bellini}  \& {Renzini}}{{Marino} et~al.}{2009}]{Marino2009}
{Marino} A.~F.,  {Milone} A.~P.,  {Piotto} G.,  {Villanova} S.,  {Bedin} L.~R.,
   {Bellini} A.,   {Renzini} A.,  2009, \mn@doi [\aap]
  {10.1051/0004-6361/200911827}, \href
  {https://ui.adsabs.harvard.edu/abs/2009A&A...505.1099M} {505, 1099}

\bibitem[\protect\citeauthoryear{{Marino} et~al.,}{{Marino}
  et~al.}{2011}]{Marino}
{Marino} et~al., 2011, \mn@doi [A\&A] {10.1051/0004-6361/201116546}, 532, A8

\bibitem[\protect\citeauthoryear{Marino et~al.,}{Marino
  et~al.}{2015}]{Marino2015}
Marino A.~F.,  et~al., 2015, \mn@doi [Monthly Notices of the Royal Astronomical
  Society] {10.1093/mnras/stv420}, 450, 815

\bibitem[\protect\citeauthoryear{{Marino} et~al.,}{{Marino}
  et~al.}{2018}]{Marino18}
{Marino} A.~F.,  et~al., 2018, \mn@doi [\apj] {10.3847/1538-4357/aabdea}, \href
  {https://ui.adsabs.harvard.edu/abs/2018ApJ...859...81M} {859, 81}

\bibitem[\protect\citeauthoryear{{Marino} et~al.,}{{Marino}
  et~al.}{2019}]{Marino2019}
{Marino} A.~F.,  et~al., 2019, \mn@doi [\mnras] {10.1093/mnras/stz1415}, \href
  {https://ui.adsabs.harvard.edu/abs/2019MNRAS.487.3815M} {487, 3815}

\bibitem[\protect\citeauthoryear{Marino et~al.,}{Marino
  et~al.}{2021}]{Marino2021}
Marino A.~F.,  et~al., 2021, \mn@doi [The Astrophysical Journal]
  {10.3847/1538-4357/ac282c}, 923, 22

\bibitem[\protect\citeauthoryear{{Masseron}, {Merle}  \& {Hawkins}}{{Masseron}
  et~al.}{2016}]{Masseron16}
{Masseron} T.,  {Merle} T.,   {Hawkins} K.,  2016, {BACCHUS: Brussels Automatic
  Code for Characterizing High accUracy Spectra} (\mn@eprint {ascl} {1605.004})

\bibitem[\protect\citeauthoryear{{Meynet}, {Ekstr\"om, S.}  \& {Maeder,
  A.}}{{Meynet} et~al.}{2006}]{Meynet2006}
{Meynet} {Ekstr\"om, S.}  {Maeder, A.} 2006, \mn@doi [A\&A]
  {10.1051/0004-6361:20053070}, 447, 623

\bibitem[\protect\citeauthoryear{{Milone} et~al.,}{{Milone}
  et~al.}{2012}]{Milone2012}
{Milone} A.~P.,  et~al., 2012, \mn@doi [\apj] {10.1088/0004-637X/744/1/58},
  \href {https://ui.adsabs.harvard.edu/abs/2012ApJ...744...58M} {744, 58}

\bibitem[\protect\citeauthoryear{{Milone} et~al.,}{{Milone}
  et~al.}{2013}]{Milone2013}
{Milone} A.~P.,  et~al., 2013, \mn@doi [\apj] {10.1088/0004-637X/767/2/120},
  \href {https://ui.adsabs.harvard.edu/abs/2013ApJ...767..120M} {767, 120}

\bibitem[\protect\citeauthoryear{{Milone} et~al.,}{{Milone}
  et~al.}{2015a}]{Milone2015a}
{Milone} A.~P.,  et~al., 2015a, \mn@doi [\mnras] {10.1093/mnras/stu2446}, \href
  {https://ui.adsabs.harvard.edu/abs/2015MNRAS.447..927M} {447, 927}

\bibitem[\protect\citeauthoryear{{Milone} et~al.,}{{Milone}
  et~al.}{2015b}]{Milone2015b}
{Milone} A.~P.,  et~al., 2015b, \mn@doi [\apj] {10.1088/0004-637X/808/1/51},
  \href {https://ui.adsabs.harvard.edu/abs/2015ApJ...808...51M} {808, 51}

\bibitem[\protect\citeauthoryear{{Milone} et~al.,}{{Milone}
  et~al.}{2017}]{Milone17}
{Milone} A.~P.,  et~al., 2017, \mn@doi [\mnras] {10.1093/mnras/stw2531}, \href
  {https://ui.adsabs.harvard.edu/abs/2017MNRAS.464.3636M} {464, 3636}

\bibitem[\protect\citeauthoryear{{Mott}, {Steffen, M.}, {Caffau, E.}  \&
  {Strassmeier, K. G.}}{{Mott} et~al.}{2020}]{Mott2020}
{Mott} {Steffen, M.} {Caffau, E.}  {Strassmeier, K. G.} 2020, \mn@doi [A\&A]
  {10.1051/0004-6361/201937047}, 638, A58

\bibitem[\protect\citeauthoryear{{Mu{\~n}oz}, {Geisler}, {Villanova},
  {Sarajedini}, {Frelijj}, {Vargas}, {Monaco}  \& {O'Connell}}{{Mu{\~n}oz}
  et~al.}{2021}]{Munoz21}
{Mu{\~n}oz} C.,  {Geisler} D.,  {Villanova} S.,  {Sarajedini} A.,  {Frelijj}
  H.,  {Vargas} C.,  {Monaco} L.,   {O'Connell} J.,  2021, arXiv e-prints,
  \href {https://ui.adsabs.harvard.edu/abs/2021arXiv210615052M} {p.
  arXiv:2106.15052}

\bibitem[\protect\citeauthoryear{Mucciarelli, Lapenna, Massari, Pancino,
  Stetson, Ferraro, Lanzoni  \& Lardo}{Mucciarelli
  et~al.}{2015}]{Mucciarelli2015}
Mucciarelli A.,  Lapenna E.,  Massari D.,  Pancino E.,  Stetson P.~B.,  Ferraro
  F.~R.,  Lanzoni B.,   Lardo C.,  2015, \mn@doi [The Astrophysical Journal]
  {10.1088/0004-637x/809/2/128}, 809, 128

\bibitem[\protect\citeauthoryear{Mészáros et~al.,}{Mészáros
  et~al.}{2020}]{Meszaros20}
Mészáros S.,  et~al., 2020, \mn@doi [Monthly Notices of the Royal
  Astronomical Society] {10.1093/mnras/stz3496}, 492, 1641–1670

\bibitem[\protect\citeauthoryear{Nardiello et~al.,}{Nardiello
  et~al.}{2018a}]{Nardiello2018}
Nardiello D.,  et~al., 2018a, \mn@doi [Monthly Notices of the Royal
  Astronomical Society] {10.1093/mnras/sty719}, 477, 2004

\bibitem[\protect\citeauthoryear{{Nardiello} et~al.,}{{Nardiello}
  et~al.}{2018b}]{2018MNRAS.481.3382N}
{Nardiello} D.,  et~al., 2018b, \mn@doi [\mnras] {10.1093/mnras/sty2515}, \href
  {https://ui.adsabs.harvard.edu/abs/2018MNRAS.481.3382N} {481, 3382}

\bibitem[\protect\citeauthoryear{{Pilachowski}, {Leep}, {Wallerstein}  \&
  {Peterson}}{{Pilachowski} et~al.}{1982}]{Pilachowski82}
{Pilachowski} C.,  {Leep} E.~M.,  {Wallerstein} G.,   {Peterson} R.~C.,  1982,
  \mn@doi [\apj] {10.1086/160493}, \href
  {https://ui.adsabs.harvard.edu/abs/1982ApJ...263..187P} {263, 187}

\bibitem[\protect\citeauthoryear{{Piotto}, {Milone}, {Marino}, {Bedin},
  {Anderson}, {Jerjen}, {Bellini}  \& {Cassisi}}{{Piotto}
  et~al.}{2013}]{Piotto2013}
{Piotto} G.,  {Milone} A.~P.,  {Marino} A.~F.,  {Bedin} L.~R.,  {Anderson} J.,
  {Jerjen} H.,  {Bellini} A.,   {Cassisi} S.,  2013, \mn@doi [\apj]
  {10.1088/0004-637X/775/1/15}, \href
  {https://ui.adsabs.harvard.edu/abs/2013ApJ...775...15P} {775, 15}

\bibitem[\protect\citeauthoryear{{Piotto} et~al.,}{{Piotto}
  et~al.}{2015}]{2015AJ....149...91P}
{Piotto} G.,  et~al., 2015, \mn@doi [\aj] {10.1088/0004-6256/149/3/91}, \href
  {https://ui.adsabs.harvard.edu/abs/2015AJ....149...91P} {149, 91}

\bibitem[\protect\citeauthoryear{Renzini et~al.,}{Renzini
  et~al.}{2015}]{Renzini}
Renzini A.,  et~al., 2015, \mn@doi [Monthly Notices of the Royal Astronomical
  Society] {10.1093/mnras/stv2268}, 454, 4197

\bibitem[\protect\citeauthoryear{{Salaris}, {Chieffi}  \&
  {Straniero}}{{Salaris} et~al.}{1993}]{1993ApJ...414..580S}
{Salaris} M.,  {Chieffi} A.,   {Straniero} O.,  1993, \mn@doi [\apj]
  {10.1086/173105}, \href
  {https://ui.adsabs.harvard.edu/abs/1993ApJ...414..580S} {414, 580}

\bibitem[\protect\citeauthoryear{{Shetrone} \& {Keane}}{{Shetrone} \&
  {Keane}}{2000}]{Shetrone2000}
{Shetrone} M.~D.,  {Keane} M.~J.,  2000, \mn@doi [\aj] {10.1086/301232}, \href
  {https://ui.adsabs.harvard.edu/abs/2000AJ....119..840S} {119, 840}

\bibitem[\protect\citeauthoryear{{Sneden}}{{Sneden}}{1973}]{1973ApJ...184..839S}
{Sneden} C.,  1973, \mn@doi [\apj] {10.1086/152374}, \href
  {https://ui.adsabs.harvard.edu/abs/1973ApJ...184..839S} {184, 839}

\bibitem[\protect\citeauthoryear{{Tautvai{\v{s}}ien{\.{e}}}
  et~al.,}{{Tautvai{\v{s}}ien{\.{e}}} et~al.}{2022}]{Tautvaisien2022}
{Tautvai{\v{s}}ien{\.{e}}} G.,  et~al., 2022, \mn@doi [\aap]
  {10.1051/0004-6361/202142234}, \href
  {https://ui.adsabs.harvard.edu/abs/2022A&A...658A..80T} {658, A80}

\bibitem[\protect\citeauthoryear{Vasiliev \& Baumgardt}{Vasiliev \&
  Baumgardt}{2021}]{Vasiliev2021}
Vasiliev E.,  Baumgardt H.,  2021, Gaia EDR3 view on Galactic globular clusters
  (\mn@eprint {arXiv} {2102.09568})

\bibitem[\protect\citeauthoryear{{Ventura}, {D'Antona}, {Mazzitelli}  \&
  {Gratton}}{{Ventura} et~al.}{2001}]{Ventura2001}
{Ventura} P.,  {D'Antona} F.,  {Mazzitelli} I.,   {Gratton} R.,  2001, \mn@doi
  [\apjl] {10.1086/319496}, \href
  {https://ui.adsabs.harvard.edu/abs/2001ApJ...550L..65V} {550, L65}

\bibitem[\protect\citeauthoryear{Villanova, Geisler, Carraro, Bidin  \&
  Muñoz}{Villanova et~al.}{2013}]{Villanova2013}
Villanova S.,  Geisler D.,  Carraro G.,  Bidin C.,   Muñoz C.,  2013, \mn@doi
  [The Astrophysical Journal] {10.1088/0004-637X/778/2/186}, 778

\bibitem[\protect\citeauthoryear{{Willman} \& {Strader}}{{Willman} \&
  {Strader}}{2012}]{Willman}
{Willman} B.,  {Strader} J.,  2012, \mn@doi [\aj] {10.1088/0004-6256/144/3/76},
  \href {https://ui.adsabs.harvard.edu/abs/2012AJ....144...76W} {144, 76}

\bibitem[\protect\citeauthoryear{Worley \& Cottrell}{Worley \&
  Cottrell}{2010}]{Worley10}
Worley C.,  Cottrell P.,  2010, \mn@doi [Monthly Notices of the Royal
  Astronomical Society] {10.1111/j.1365-2966.2010.16837.x}, 406, 2504

\bibitem[\protect\citeauthoryear{Yong et~al.,}{Yong et~al.}{2014}]{Yong2014}
Yong D.,  et~al., 2014, \mn@doi [Monthly Notices of the Royal Astronomical
  Society] {10.1093/mnras/stu806}, 441, 3396

\bibitem[\protect\citeauthoryear{de Mink, Pols, Langer  \& Izzard}{de~Mink
  et~al.}{2009}]{deMink:2009yg}
de Mink S.~E.,  Pols O.~R.,  Langer N.,   Izzard R.~G.,  2009, \mn@doi [IAU
  Symp.] {10.1017/S1743921309991025}, 266, 169

\makeatother
\end{thebibliography}

\bsp	
\label{lastpage}
\end{document}